\documentclass[11pt]{article}
\pdfoutput=1
\usepackage{amsmath}
\usepackage{amssymb}
\usepackage{multirow}

\usepackage{graphicx,bbm,mathrsfs}
\usepackage{nicefrac}
\usepackage{bbm}
\usepackage{geometry}       
\usepackage{jheppub}     
\usepackage{bm}        
\usepackage{graphicx,mathrsfs}
\usepackage{nicefrac}
\usepackage{color}
\definecolor{orange}{RGB}{ 148,0,211}

\def\ie{{\it i.e.}}

\newcommand{\be}{\begin{equation}}  
\newcommand{\ee}{\end{equation}}  
\newcommand{\bea}{\begin{eqnarray}}  
\newcommand{\eea}{\end{eqnarray}}  

\newcommand{\tr}{\operatorname{tr}}

\newcommand{\bmat}{\begin{pmatrix}}
\newcommand{\emat}{\end{pmatrix}}

\newcommand*{\Scale}[2][4]{\scalebox{#1}{$#2$}}%
\begin{document}
                
\title{The Global Higgs as a Signal for Compositeness at the LHC}

\author{Sylvain~Fichet$\,^a$,}
\emailAdd{sylvain.fichet@gmail.com}

\author{Gero von Gersdorff$\, ^b$,}
\emailAdd{gersdorff@gmail.com}

\author{Eduardo Pont\'on$\,^a$ and}
\emailAdd{eponton@ift.unesp.br}

\author{Rogerio Rosenfeld$\,^a$}
\emailAdd{rosenfel@ift.unesp.br}

\affiliation{$^a$ ICTP South American Institute for Fundamental
  Research \& Instituto de F\'isica Te\'orica \\ 
  Universidade Estadual  Paulista, S\~ao Paulo, Brazil}

\affiliation{$^b$ Departamento de F\'isica, Pontif\'icia Universidade
Cat\'olica de Rio de Janeiro, Rio de Janeiro, Brazil}

\abstract{ The radial excitation of the global symmetry-breaking
vacuum in composite Higgs models, called the ``global Higgs'', has
been recently a focus of investigation.  In this paper we study the
prospects for detecting this composite scalar at the 13 TeV LHC. We compute the global Higgs
production rates and estimate the discovery potential of a global
Higgs decaying into top quark pairs and into Higgs and electroweak
gauge bosons with subsequent hadronic decays.  The global Higgs may
also decay into fermion resonances such as top partners, providing a
new window into compositeness.  We show that top partner jets can be
effectively unresolved in some regions of the parameter space.  Such
``boosted top partner'' signatures would deserve the development of
dedicated substructure analyses.  }

{\center \today}

\maketitle

\thispagestyle{empty}
\setcounter{page}{1}

\section{Introduction}
\label{se:intro}

The existence of another spin-0, CP-even particle in viable composite
Higgs models where the Higgs field is identified as a pseudo
Nambu-Goldstone boson (pNGB) has recently been a topic of
investigation in the literature \cite{Buttazzo:2015bka,Feruglio:2016zvt,us}.  This
heavy state was defined in~\cite{us} as the ``radial" excitation of
the coset space parameterized by the NGBs of such models, and is
referred to as the \textit{global Higgs}.  It is therefore intimately
connected to the breaking of an (approximate) global symmetry in a new
strongly-coupled sector, and identifying it would give information
about this breaking, equivalent to the information we have obtained
about electroweak symmetry breaking (EWSB) from the observation of the
Higgs resonance at the LHC~\cite{Aad:2012tfa,Chatrchyan:2012xdj}.

It has been shown in \cite{vonGersdorff:2015fta,us} that the global
Higgs can consistently be amongst the lightest states of the strongly
coupled theory, probably around the scale of the lowest lying
fermionic resonances, and below the scale of spin-1 excitations.
Moreover, it couples in a model-independent manner to the (SM) Higgs
boson and to the longitudinal electroweak gauge bosons, with a
sizeable strength.  Interestingly, its one-loop interactions with
transverse gauge bosons (in particular its couplings to gluons) can be
enhanced by the large number of states running in the loops.

In this work we focus on the LHC implications of such a global Higgs
particle, which could very well be the first signal of Higgs
compositeness at the LHC.

Our plan is as follows.  The properties of the global Higgs that are
relevant for the LHC phenomenology are summarized in
Sec.~\ref{se:coup}, and its main production rates are evaluated in
Sec.~\ref{se:production}.  The discovery potential of the global Higgs
at the $13$ TeV LHC is estimated in Sec.~\ref{se:diboson}, through its
decays into NGBs and top pairs.  In Sec.~\ref{se:top_partners}, we
consider the case of top partner resonant production via the global
Higgs channel, and the possibility of boosted top partner signals is
subsequently investigated.  We conclude in Sec.~\ref{se:conclusions}.

\section{Properties of the Global Higgs}
\label{se:coup}

The properties of the global Higgs were presented in Ref.~\cite{us}
for the case of the $SO(5)/SO(4)$ coset.  For concreteness, we
continue focusing on this example.  In this section, we summarize the
features that are relevant for studying the LHC phenomenology of the
global Higgs.

\subsection{Tree-level couplings}

The mass, vacuum expectation value (VEV) and quartic coupling of the
global Higgs are denoted by $m_\phi$, $\hat f$ and $\lambda$,
respectively.  They are related by
\be m_\phi=\sqrt{2\lambda} \, {\hat f}~.
\ee 
The NGBs of the $SO(5) \to SO(4)$ breaking, which belong to the same
$SO(5)$ multiplet as the global Higgs, can mix with the longitudinal
components of massive spin-1 resonances of the underlying strong
dynamics.  As a result, their decay constant $f$ (which controls the
deviations of the pNGB Higgs from the SM limit) is expected to be
smaller than $\hat f$ (which controls the couplings of the global
Higgs).  This extra degree of freedom is parameterized by
\be
r_v=\frac{f^2}{\hat f ^2}\,, \quad \textrm{where} \quad r_v \leq 1~.
\ee
Among the SM particles, the global Higgs couples mainly to the
$SO(5)/SO(4)$ NGBs, i.e.~to the Higgs boson and the longitudinal
polarizations of the $W$ and $Z$, and to the top quark.  The
corresponding couplings are read from
\bea
{\cal L} &\supset& 2 \frac{ r_v}{\hat f} \, \phi\, |D_\mu H|^2 - \frac{m_t}{\hat f} \phi\, \bar t t~.
\eea
These depend on two independent parameters, $\hat{f}$ and $r_v$, and
lead to the 2-body decay widths (neglecting EWSB effects):
\be
\Gamma_{\phi\rightarrow h h} ~=~\Gamma_{\phi\rightarrow Z_LZ_L } ~=~ \frac{1}{2} \, \Gamma_{\phi\rightarrow W_L^+W_L^-} ~=~ \frac{r_v^2}{32\pi}\frac{m_\phi^3}{\hat f^2}~,
\label{Gammahh}
\ee
and
\be
\Gamma_{\phi\rightarrow t \bar t } = N_c\frac{m_t^2}{8\pi \hat f ^2}\, m_\phi~.
\label{Gammatt}
\ee

The global Higgs also couples to the heavy fermion and vector
resonances of the theory.  The vector resonances affect the global
Higgs phenomenology mainly at loop level, to be discussed in the next
subsection.  Regarding the spin-$1/2$ resonances, however, one should keep
in mind that the global Higgs can have a sizeable branching fraction
into a fermion resonance plus a SM fermion $\phi \rightarrow \psi_{\rm
SM}\bar\psi$, $\phi \rightarrow \psi \bar\psi_{\rm SM}$, or into two
heavy fermion resonances $\phi\rightarrow \psi \bar\psi$, provided
such decays are kinematically allowed.  The precise branching
fractions are highly model-dependent, but when open such channels
typically dominate the decay modes of the global Higgs.

\subsection{Loop-induced couplings}

The heavy resonances induce additional couplings of the global Higgs
to the SM gauge fields.  These can be parameterized by local operators
as follows:
\be
{\cal L} \supset -\phi\left( \frac{ a_{ gg}}{ \hat{f}}  \, (G^a_{\mu\nu} )^2
+  \frac{a_{ WW}}{ \hat{f}}  \, W^+_{\mu\nu} W^{-\mu\nu }
+  \frac{a_{ ZZ}}{ \hat{f}}  \, (Z_{\mu\nu})^2 
+  \frac{a_{ \gamma \gamma}}{ \hat{f}} \,  (F_{\mu\nu} )^2
+  \frac{a_{ \gamma Z}}{ \hat{f}}  \, F_{\mu\nu} Z^{\mu\nu} \right)~,
\label{Lphieff}
\ee
where the coefficients $a_i$ depend on the detailed spectrum of heavy
resonances.  The most important of these, from a phenomenological
point of view, is the coupling to gluons which plays a crucial role in
the production of the global Higgs at the LHC. The above include also
couplings to two photons and to the transverse degrees of freedom of
the $W$ and $Z$ vector bosons, which can play a non-negligible role in
some regions of the parameter space.

The contribution to the $a_i$ due to vector resonances depends only on
the $SO(5)/SO(4)$ coset structure.  We expect these spin-1 resonances
to be heavy compared to the global Higgs, which implies that the
corresponding contribution depends only on $r_v$~\cite{us}.

The fermion contribution, on the other hand, depends on the specifics
of the fermionic sector.  Instead of trying to perform a detailed
analysis by scanning over the full set of microscopic parameters of
given models, we will establish a reasonable ``model-independent"
estimate that captures the expected size of the $a_i$'s within a
factor of order one, at least in the bulk of the natural parameter
space of the models we envision (see later).  This will also allow us
to explore the potential enhancements due to the multiplicity of
resonances that get part of their mass from the breaking of the global
symmetry.\footnote{Each SM fermion can have an associated tower of
resonances.  We focus on the ``first level" of resonances, as would
arise, for instance, in a two-site construction.  These have a
multiplicity dictated by the $SO(4)$ representations they belong to.
Their masses are split only due to mixing with the ``elementary"
fermion sector.  In addition, several $SO(4)$ multiplets can fit into
$SO(5)$ multiplets, which we assume receive a common ``vectorlike"
mass, i.e.~independent of the global symmetry breaking scale
$\hat{f}$.  The mass splitting of the various $SO(4)$ multiplets
belonging to the same $SO(5)$ multiplet is controlled by the scale of
global symmetry breaking, and by the strength of the Yukawa
interactions coupling the global Higgs to the fermion resonances.
See~\cite{us} for a more detailed discussion.  We also note that there
can be heavier resonances (belonging to a ``second" or higher levels),
which are expected to give a subdominant contribution to the
loop-induced couplings.} Following Ref.~\cite{us}, we first note that
the $a_i$'s can be usefully thought as containing two distinct
ingredients.  First, the 1-loop integral itself depends only on the
physical fermion masses, $M_i$, and on the global Higgs mass, through
the combination $\tau_i = m^2_\phi / (4 M^2_i)$.  When the $M_i$ are
of order, or larger than $m_\phi$, the (dimensionless) loop-function,
commonly denoted by $A_{1/2}(\tau_i)$ and given in
App.~\ref{loopfunctions}, displays a mild dependence on $M_i$.  For
instance, when $0.6 \, m_\phi \lesssim M_i < \infty$, the loop
function $A_{1/2}(\tau_i)$ deviates by at most $20\%$ from its value
at $M_i = m_\phi$.  For the cases of interest in this work, we can
parametrize the scale of heavy fermionic resonances by a single
``average" mass scale that we denote by $\bar{M}_\psi$, and take the
loop function as (approximately) universal: $A_{1/2}(\tau_i) \approx
A_{1/2}(m^2_\phi / (4 \bar{M}^2_\psi))$.\footnote{If there is some
resonance significantly lighter than the global Higgs, the above
overestimates the loop function.  In such a scenario, the global Higgs
will decay dominantly into the fermionic channel, a case we will treat
separately in this work.} The second, more important ingredient, is
the actual coupling of any given fermionic resonance to the global
Higgs.  Such a coupling depends on the underlying (proto-)Yukawa
coupling and on an angle that characterizes the mixing between
different $SO(4)$ representations.  The mixing angle parametrizes the
fraction of the fermion mass coming from the breaking of the global
symmetry, and therefore describes the decoupling properties of these
virtual effects.  Under the assumption discussed above of a universal
loop function, we can use well known sum rules to define a convenient
reference fermion scale, $M_\psi$.  For example, for the gluon fusion
process, we write
\bea
\hat{f} \, \sum_i \frac{M_i'}{M_i} = -2\frac{\hat{f}^2}{M_\psi^2}\left(\bar N^U_{\phi g g}\tr \xi_U'\xi_U^T +\bar N^D_{\phi gg }\tr \xi'_D\xi_D^T \right)~,
\label{Mpsi}
\eea
where $M_i' = dM_i/d\hat{f}$ and the proto-Yukawa couplings,
$\xi^{(\prime)}_U$ and $\xi^{(\prime)}_D$, were defined in~\cite{us}.
There are similar expressions for $\phi BB$ and $\phi \gamma\gamma$,
where the sums are now weighted by the square of the hypercharges and
charges, respectively (see Ref.~\cite{us} for the explicit
expressions).  The trace is over generations, and the $N^{U,D}_A$,
with $A = \phi g g$, $\phi BB$ and $\phi \gamma\gamma$ characterize
the multiplicity effect of a given tower of resonances associated with
the up or down sectors.  Analogous leptonic multiplicities, $N^{E}_A$,
enter into the $\phi BB$ and $\phi \gamma\gamma$ processes.  Since
$M_\psi$ and $\bar{M}_\psi$ are typically close, we simply take
$\bar{M}_\psi = M_\psi$ in the fermion loop functions, where in
practice we think of $M_\psi$ as being defined by the gluon fusion
process.\footnote{We have checked that the scales defined from the
$\phi BB$ and $\phi \gamma\gamma$ processes are typically close to
$M_\psi$, so that they can be replaced by $M_\psi$ within the
precisions we can expect in our simplified analysis.  In other words,
as far as the loop processes are concerned, in a large region of
parameter space, the fermion sector can be characterized by a
\textit{single} scale of fermionic resonances, $M_\psi$, and by
multiplicities that depend only on the field content and quantum
numbers, but not on the parameters of the model.}

Putting the previous ingredients together, we write
\bea
a_{ gg} =  - c_{ gg} \, \frac{\hat f^2}{M_\psi^2} \, A_{1/2}\!\left(\frac{m_\phi^2}{4M_\psi^2}\right)~,
\quad a_{ BB} =  - c_{ BB} \, \frac{\hat f^2}{M_\psi^2} \, A_{1/2}\!\left(\frac{m_\phi^2}{4M_\psi^2}\right)~,
\label{cphigg}
\eea
\be
a_{ \gamma\gamma}=  - c_{ \gamma\gamma} \, \frac{\hat f^2}{M_\psi^2} \, A_{1/2}\!\left(\frac{m_\phi^2}{4M_\psi^2}\right) - 0.0022 \, (1-r_v)\,,
\label{cphiBB}
\ee
together with the $SU(2)_L \times U(1)_Y$ relations
\be
a_{ WW} = \frac{2}{s^2_W}(a_{ \gamma\gamma} - c^2_W a_{ BB})~,~
a_{ ZZ} = \frac{1}{2} \, c^2_Wa_{ WW}+s^2_Wa_{ BB}~,~
a_{ \gamma Z } = s_W c_W(a_{ WW}-2a_{ BB})~.
\ee
In Eq.~\eqref{cphiBB} we used that $(\alpha/8\pi) A_1\approx 0.0022$ in the asymptotic limit where $A_1\rightarrow -7$.
The $c_i$ coefficients are given by
\bea
c_{gg} &=& \frac{\alpha_s}{8\pi} \left(\bar N^U_{\phi g g}\tr \xi_U'\xi_U^T +\bar N^D_{\phi gg }\tr \xi'_D\xi_D^T \right)~,
\nonumber \\[0.4em]
c_{BB} &=& \frac{\alpha}{4\pi c^2_W} \left(N_c \, \bar N^U_{\phi B B}\tr \xi_U'\xi_U^T + N_c \, \bar N^D_{\phi B B}\tr \xi'_D\xi_D^T + \bar N^E_{\phi B B}\tr \xi_E'\xi_E^T \right)~,
\label{cs} \\[0.4em]
c_{\gamma\gamma} &=& \frac{\alpha}{4\pi} \left(N_c \, \bar N^U_{\phi \gamma\gamma}\tr \xi_U'\xi_U^T + N_c \, \bar N^D_{\phi \gamma\gamma}\tr \xi'_D\xi_D^T + \bar N^E_{\phi \gamma\gamma}\tr \xi_E'\xi_E^T \right)~,
\nonumber
\eea
where $N_c = 3$ is the number of colors, $\alpha_s$ is the strong
coupling constant, $\alpha$ is the fine structure constant, and the
multiplicities, $N^{U,D,E}_A$, with $A = \phi g g$, $\phi BB$ and
$\phi \gamma\gamma$ encode the model-dependence (to be discussed
next).

The partial decay widths into transverse gauge bosons are given by
\be
\Gamma_{\phi\rightarrow g g} = \frac{2a_{gg}^2}{\pi}\,\frac{m_\phi^3}{\hat f^2}~,
\ee
\be
\Gamma_{\phi\rightarrow \gamma\gamma} = \frac{a_{\gamma\gamma}^2}{4\pi}\,\frac{m_\phi^3}{\hat f^2}~,\quad \quad
\Gamma_{\phi\rightarrow Z_TZ_T} = \frac{a_{ZZ}^2}{4\pi}\,\frac{m_\phi^3}{\hat f^2}~,
\ee
\be
\Gamma_{\phi\rightarrow \gamma Z_T} = \frac{a_{\gamma Z}^2}{8\pi}\,\frac{m_\phi^3}{\hat f^2}~,\quad\quad
\Gamma_{\phi\rightarrow W_T^+W_T^-}= \frac{a_{WW}^2}{8\pi}\,\frac{m_\phi^3}{\hat f^2}~.
\ee

\subsection{Benchmark scenarios}

In Ref.~\cite{us}, we defined a number of fermion realizations, which
differ by the $SO(5)$ embeddings of the fermion partners and of the
global Higgs.  These benchmark scenarios are defined by
\begin{table}[h]
\begin{center}
\begin{tabular}{lllc}
$\bullet$~~MCHM$_{5,1,10}$: 
& $(Q_i, U_i, D_i) = ({\bf 5}_{\frac{2}{3}}, {\bf 1}_{\frac{2}{3}}, {\bf 10}_{\frac{2}{3}})$~,
& $\phi \subset {\bf 5_0}$~,
& 
\\ [0.5em]
$\bullet$~~MCHM$_{5,14,10}$: 
& $(Q_i, U_i, D_i) = ({\bf 5}_{\frac{2}{3}}, {\bf 14}_{\frac{2}{3}}, {\bf 10}_{\frac{2}{3}})$~,
& $\phi \subset {\bf 5_0}$~,
\\ [0.5em]
$\bullet$~~MCHM$_{14,14,10}$: 
& $(Q_i, U_i, D_i) = ({\bf 14}_{\frac{2}{3}}, {\bf 14}_{\frac{2}{3}}, {\bf 10}_{\frac{2}{3}})$~, 
& $\phi \subset {\bf 14_0}$~,
\\ [0.5em]
$\bullet$~~MCHM$_{5,1}$: 
& $(Q_3, U_3) = ({\bf 5}_{\frac{2}{3}}, {\bf 1}_{\frac{2}{3}})$~, 
& $\phi \subset {\bf 5_0}$~,
\end{tabular}
\end{center}
\end{table}

\vspace{-5mm} 
\noindent 
where we indicate the $SO(5)$ representations and $U(1)_X$ charges of
the ``partners" of the SM $SU(2)_L$ quark doublets and up-type quark
and down-type quark singlets, as well as of the global Higgs
multiplet.  The precise embedding of the lepton sector affects the
electroweak channels, such as $\phi \to \gamma\gamma$, $\phi \to
\gamma Z_T$ and $\phi \to V_T V_T$, and we refer the reader to
Ref.~\cite{us} for illustrative benchmarks.  We will use the
multiplicities $\bar N^{U,D,E}_{\phi\gamma\gamma}$ and $\bar
N^{U,D,E}_{\phi BB}$ computed in that reference, and reproduced in
Table~\ref{tab:multi}.  The last model defined above, MCHM$_{5,1}$, is
a ``non-anarchic" scenario where only the top quark resonances give a
non-negligible effect.  Further details can be found in~\cite{us}.

\begin{table}[t]
\begin{center}
\begin{tabular}{|c||cc|ccc|ccc|}
\hline 
\rule{0mm}{5mm}
Benchmark	& $\bar{N}^U_{\phi gg}$ 			& $\bar{N}^D_{\phi gg}$ 	
			& $\bar{N}^U_{\phi \gamma\gamma}$	& $\bar{N}^D_{\phi \gamma\gamma}$ & $\bar{N}^E_{\phi \gamma\gamma}$ 	
			& $\bar{N}^U_{\phi BB}$ 			& $\bar{N}^D_{\phi BB}$ & $\bar{N}^E_{\phi BB}$ 
\\ [0.1em]
\hline 
\hline 
\rule{0mm}{5mm}
MCHM$_{5,1,10}$	& 1	&2	&$\frac{4}{9}$	& $\frac{17}{9}$ &		1	& $\frac{4}{9}$ & $\frac{25}{18}$ &1
\\ [0.4em]
\hline
\rule{0mm}{5mm}
MCHM$_{5,14,10}$& $\frac{14}{5}$	& 2	& $\frac{101}{45}$	& $\frac{17}{9}$			&1	& $\frac{157}{90}$	& $\frac{25}{18}$ &1
\\ [0.4em]
\hline
\rule{0mm}{5mm}
MCHM$_{14,14,10}$&$\frac{27}{20}$	& $\frac{5}{4}$	&	 $\frac{57}{40}$	&$\frac{85}{72}$ &	1		& $\frac{81}{80}$ &$\frac{125}{144}$ &1
\\ [0.4em]
\hline 
\rule{0mm}{5mm}
MCHM$_{5,1}$	& 1	&$-$	&$\frac{4}{9}$	& $-$ & $-$	& $\frac{4}{9}$ & $-$ & $-$
\\ [0.4em]
\hline 
\end{tabular}
\caption{ Fermionic multiplicity factors entering the effective
couplings of the global Higgs to two gluons or two EW gauge bosons.
Reproduced from Ref~\cite{us}.
\label{tab:multi}}
\end{center}
\end{table}

We will make the reasonable assumption that the vector-like masses are
of the same order for all the resonances.  We then note that when the
global symmetry breaking effects are small compared to such
vector-like masses, and when the mixing between the elementary and
composite sectors is small (as may be expected for the quarks other
than the top quark), the scale $M_\psi$ in Eq.~(\ref{Mpsi}) coincides
with the ``universal" vector-like mass.  When either the elementary
composite mixing is large (as would be the case for the top sector) or
if the global symmetry breaking contributions to the fermions masses
are sizeable, the scale $M_\Psi$ can differ by an order one factor
from the vector-like parameters.  Typically, however, this scale is of
the same order as the physical fermion masses and, as described above,
we incur in small errors if we identify $M_\psi$ [defined by
Eq.~(\ref{Mpsi})] with the average fermion mass used in the loop
function.

In reference~\cite{us} we also estimated for each benchmark scenario
the expected size of the proto-Yukawa couplings by assuming that they
are all of the same order (we call it $\xi$) and requiring
perturbativity up to a scale a few times above $m_\phi$.  This results
in
\bea
\xi &\approx& 0.6 \hspace{5mm} \textrm{for the MCHM$_{5,1,10}$}~,
\hspace{1.3cm}
\xi ~\approx~ 0.5 \hspace{5mm} \textrm{for the MCHM$_{5,14,10}$}~,
\nonumber \\ [0.4em]
\xi &\approx& 0.6 \hspace{5mm} \textrm{for the MCHM$_{14,14,10}$}~,
\hspace{1 cm}
\xi ~\approx~ 1.6 \hspace{5mm} \textrm{for the MCHM$_{5,1}$}~,
\nonumber 
\eea
with a mild dependence on the cutoff scale.  Using this information,
and the multiplicities quoted in Table~\ref{tab:multi}, we find from
Eqs.~(\ref{cs}):~\footnote{We note that by choosing $\xi'_i =
\xi_i\,(= \xi)$ in Eqs.~(\ref{cs}), the spin-1 and spin-$1/2$
contributions add up constructively in $a_{\gamma\gamma}$.  They would
interfere destructively if the $\xi'_i$ had an opposite sign to the
$\xi_i$.  Similarly, depending on relative phases, the fermion
contributions can interfere destructively with each other.  Our
numerical choice then corresponds to an optimistic scenario.}
\be
 c_{ gg}=\begin{pmatrix}
 0.013 \\0.014 \\0.011 \\ 0.010 
 \end{pmatrix}\,, \quad  c_{ BB}=\begin{pmatrix}
 0.0057 \\0.0063 \\0.0058 \\ 0.0028 
 \end{pmatrix} \,,\quad
 c_{ \gamma\gamma}=\begin{pmatrix}
 0.0054 \\0.0063 \\0.0060 \\ 0.0021 
 \end{pmatrix}\,,
\label{cphi}
\ee
where the four lines correspond to the four benchmarks defined above.
We used here $\alpha_s = 0.1$, $\alpha = 1/127$ and $s^2_W = 0.231$.

The above set of benchmark models was chosen to exhibit a broad range
of multiplicities of fermionic resonances.  We see, however, that the
above coefficients are nearly model independent.\footnote{Only
$c_{BB}$ and $c_{\gamma\gamma}$ in the MCHM$_{5,1}$ differ by a factor
of $2-3$ from the other ``high-multiplicity" models.} The reason is
that the same multiplicity factors entering in the triangle diagram
also enter in the dominant contribution to the $\beta$-functions of
the proto-Yukawa couplings.  The enhancement due to the number of
states is then largely compensated by the requirement to take a
smaller proto-Yukawa coupling (at the scale of $m_\phi$), or else a
Landau pole will develop too close to the scales of interest.  Since
the most important process for the global Higgs phenomenology is the
gluon fusion process, we will simply take, based on the above
findings, $c_{gg} \approx 0.01$ in our phenomenological study.  We
will, however, include a $K$-factor of $K \approx 2$~\cite{Djouadi:2005gi}.

\subsection{Parameter space}

We set the decay constant of the NGBs, $f$, to its approximate
experimental lower bound~\cite{Panico:2015jxa}
\be 
f=800~{\rm GeV} ~.
\ee
This ensures that the (SM) Higgs sector is roughly consistent with the
present Higgs constraints, while minimizing the fine-tuning of the
electroweak scale.  As discussed above, in the bulk of the parameter
space of the scenarios considered, the global Higgs properties depend,
to a good approximation, on three real-valued parameters that can be
chosen as $m_\phi$, $\lambda$ and the ``scale of spin-$1/2$
resonances", $M_\psi$.  The other parameters defined above are
obtained via $\hat f = m_\phi/\sqrt{2\lambda} $ and $r_v=f^2/{\hat
f}^2$.  One should also remember that
\be
\hat f \geq f~.
\ee

Also, the same type of argument based on RG running that was used to
constrain the proto-Yukawa couplings $\xi$ can be used to determine a
range for the global Higgs quartic coupling.  Although the range is
model-dependent, as described in~\cite{us}, it will be sufficient to
take $\lambda \in [0.2,3]$, which falls in the correct ballpark for
the benchmark models defined above.\footnote{Such a determination is
only meant as a guide, and one cannot claim a precision beyond order
one factors.}

It is useful to note here that the loop-level couplings scale like
$\hat{f}^2/M_\psi^2 \sim (m_\phi^2/M_\psi^2) \times \lambda^{-1}$.
Therefore, they become more important for smaller $\lambda$.  On the
other hand, the tree-level couplings scale like $1/\hat{f}^{2n} \sim
\lambda^n/m_\phi^{2n}$ for a positive power, $n$.  Therefore, they
become more important for larger $\lambda$.  This competition will be
reflected in our later results.

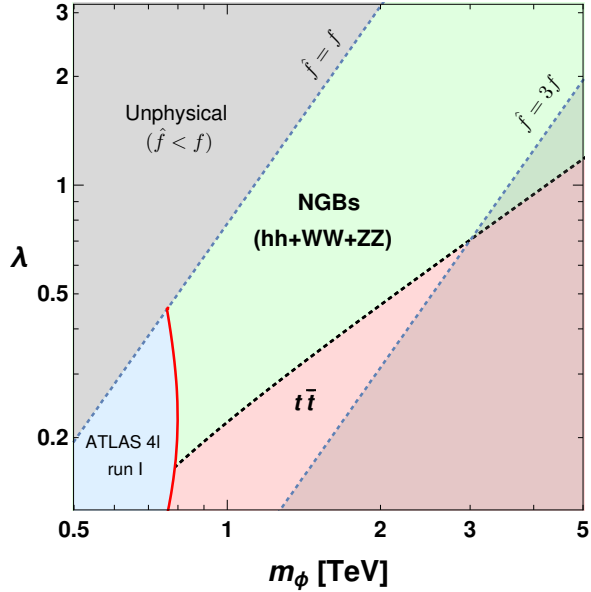
\begin{figure}
\centering
\begin{picture}(220,220)
\put(0,0){
\put(0,0){\includegraphics[scale=0.6,clip=true, trim= 0cm 0cm 0cm 0cm]{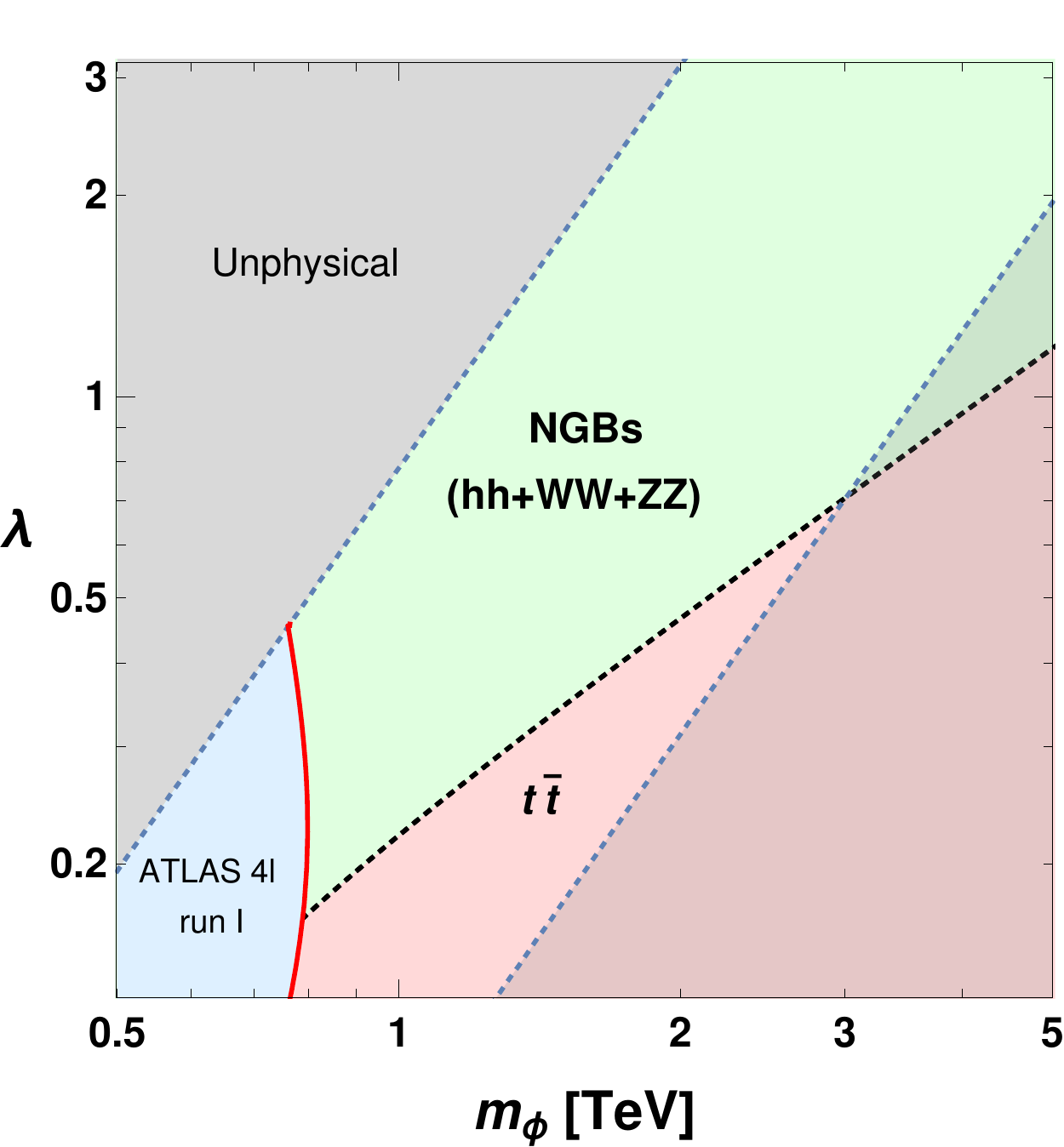}}
\put(47,167){\rotatebox{0}{ $ \Scale[0.7]{({\hat f}<f)}$}}
\put(105,190){\rotatebox{55}{ $ \Scale[0.7]{{\hat f}=f}$}}
\put(185,170){\rotatebox{55}{ $ \Scale[0.7]{{\hat f}=3f}$}}
}
\end{picture}
\caption{Regions in the $m_\phi-\lambda$ plane where the global Higgs
decays dominantly into NGBs or $t\bar{t}$ pairs, assuming that all
fermion resonances are heavier than the global Higgs.  We take $f=800$
GeV. The shaded region below the $\hat{f} =3 f$ line requires a large
hierarchy between $\hat{f}$ and $f$, and may not be realized in
typical strongly coupled scenarios.  We also show a current bound
adapted from the ATLAS heavy Higgs search of
Ref.~\cite{ATLAS-CONF-2013-013}, which shows that the global Higgs
must be heavier than about $750$~GeV.
 \label{fig:BRzones}
}
\end{figure}

\subsection{2-parameter case}

Before we undertake a study of global Higgs production, we can
immediately exhibit the relative importance of the decay channels of
the global Higgs when the fermion resonances, $\psi$, are too heavy
for any of the decays $\phi \rightarrow \psi_{\rm SM}\bar\psi$, $\phi
\rightarrow \psi \bar\psi_{\rm SM}$, or $\phi\rightarrow \psi
\bar\psi$ to be open.  The decays are then dominated by the $WW$,
$ZZ$, $hh$ and $t\bar{t}$ channels, as dictated by
Eqs.~(\ref{Gammahh}) and (\ref{Gammatt}), since the loop-induced
processes are always subdominant.  As usual, in the region where the
equivalence theorem applies, one has that the decays into $WW$, $ZZ$
and $hh$ are in the proportion $2:1:1$.  However, since these partial
widths scale like $r_v^2 m_\phi^3 / \hat{f}^2 \sim \lambda^3
f^4/m_\phi^3$, while the partial decay width into top pairs scales
like $m_t^2 m_\phi^2/\hat{f}^2 \sim m_t^2 \lambda/m_\phi$, we see that
there is a non-trivial dependence in the $m_\phi-\lambda$ plane.  The
branching fractions into NGB's and $t\bar{t}$ become equal when
$\lambda = (\sqrt{3}/2) m_t m_\phi/f^2$.  In Fig.~\ref{fig:BRzones} we
show in green the region dominated by the decays into NGBs, and in red
the region dominated by decays into top pairs.  We mark in gray the
forbidden region where $\hat{f} < f$, and also show for reference the
line where $\hat{f} =3 f$ to indicate that typically one would not
expect a large hierarchy between $\hat{f}$ and $f$.  In any case, we
see that the natural region of parameter space allows for a large
range of possibilities, although if the global Higgs is on the heavy
side of the shown range perhaps one should expect its decays to be
dominated by the NGB channels.

We also note that in the case where the decays into fermion resonances
are closed, the decay width of the global Higgs is at most
$\Gamma_{\rm tot}/m_\phi=O(0.1)$, so that the narrow width
approximation roughly applies.  If decay channels involving the
fermion resonances were open -- either mixed SM - resonance final
states or a pair of resonances -- these channels can dominate and the
global Higgs becomes a broad resonance that can reach $\Gamma_{\rm
tot}/m_\phi=O(1)$~\cite{us}.

\section{The Global Higgs at the LHC} 
\label{se:production}

The production modes of the global Higgs at the LHC have some
similarities to those of the Standard Model Higgs boson.  We focus on
inclusive resonant production
\be 
pp  \rightarrow \phi^* + X  \rightarrow Y + X~,
\ee
where $\phi^*$ means that the intermediate $\phi$ can be off-shell,
$Y$ represents the global Higgs decay products and $X$ denotes other
final states resulting of the proton collision.  Similarly to the SM
Higgs, the global Higgs can be produced through gluon fusion (ggF),
vector boson fusion (VBF), associated production with a vector boson,
and in association with a $t\bar{t}$ pair.  In principle, it could
also be produced in association with other fermion resonances, but
such production modes would be highly suppressed due to the large
masses involved.  The most important production modes are ggF and VBF,
so we focus on these two cases.

Although, as already mentioned, the global Higgs can be either a
narrow or broad resonance, typically with $\Gamma_{\rm tot}/m_\phi$
ranging from $O(10^{-3})$ to $O(1)$, we restrict here to the narrow
resonance case.  This will be sufficient for a detailed study of
scenarios where decays involving heavy resonances are closed.  Our
later remarks for cases where some such channels are open will be
treated separately.

In a large region of parameter space, the global Higgs production is
dominated by the gluon fusion process
\be 
gg  \rightarrow \phi~,
\ee
controlled by the loop-induced effects discussed in the previous
section.  As explained there, this introduces one additional parameter
beyond $m_\phi$ and $\lambda$: the scale of fermionic resonances,
$M_\psi$.  Recall that the loop-induced couplings scale like
$1/\lambda$ and therefore become larger for smaller $\lambda$.

The VBF production mode
\be 
qq'  \rightarrow \phi+qq'~,
\ee 
can proceed through tree-level couplings, which scale with $\lambda$
like $\lambda^3$, so that they can become important for larger
$\lambda$.  Note also that, as a function of $m_\phi$, these couplings
scale like $1/m^3_\phi$, for fixed $f$ and $\lambda$, and therefore
decrease quickly for a heavier global Higgs.  There are also
loop-level couplings (to transverse vector bosons and photon pairs)
that scale like $1/\lambda$ and can become important at smaller
$\lambda$.

\begin{figure}
\centering
\includegraphics[scale=0.6,clip=true, trim= 0cm 0cm 0cm 0cm]{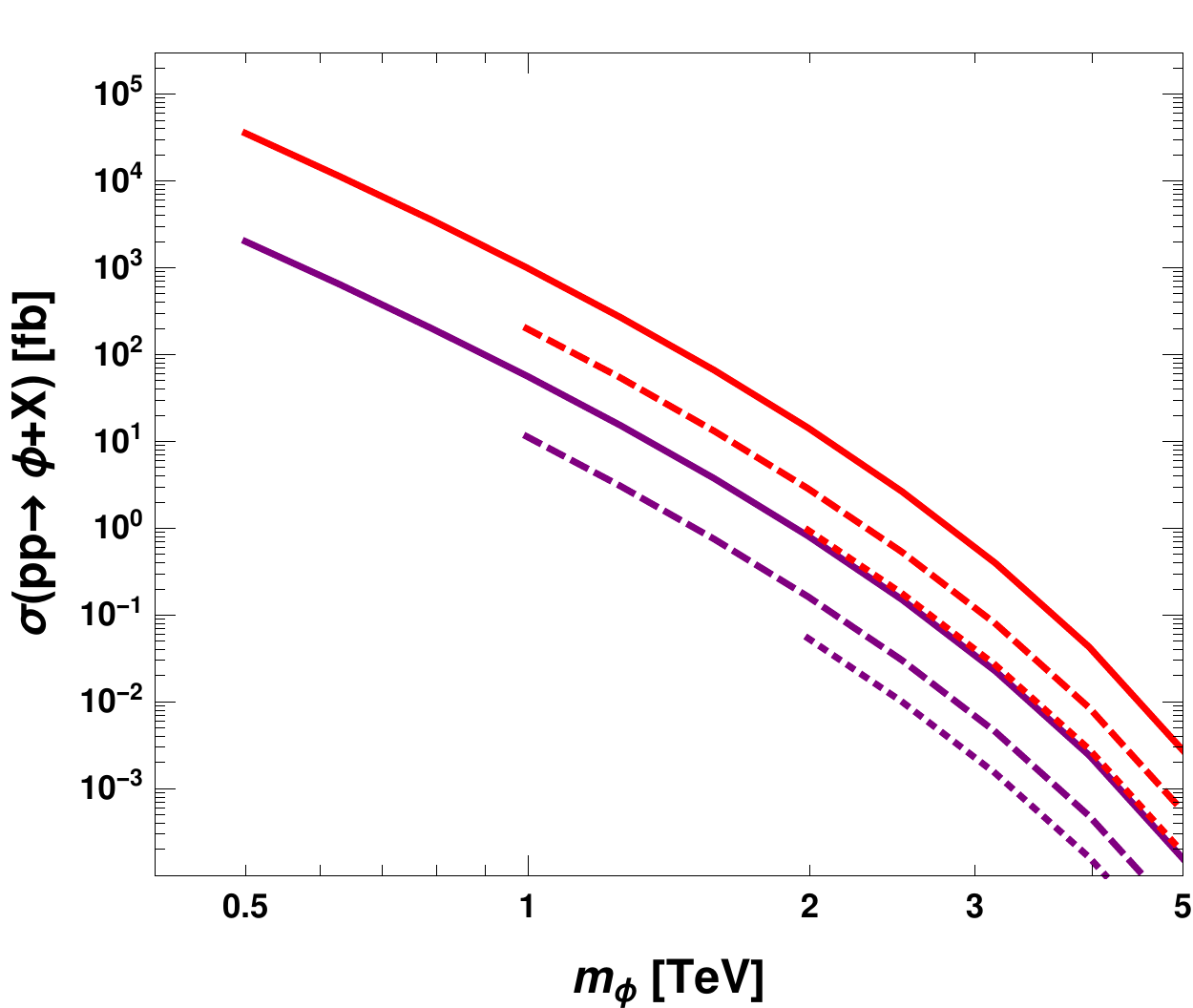}~
\includegraphics[scale=0.6,clip=true, trim= 0cm 0cm 0cm 0cm]{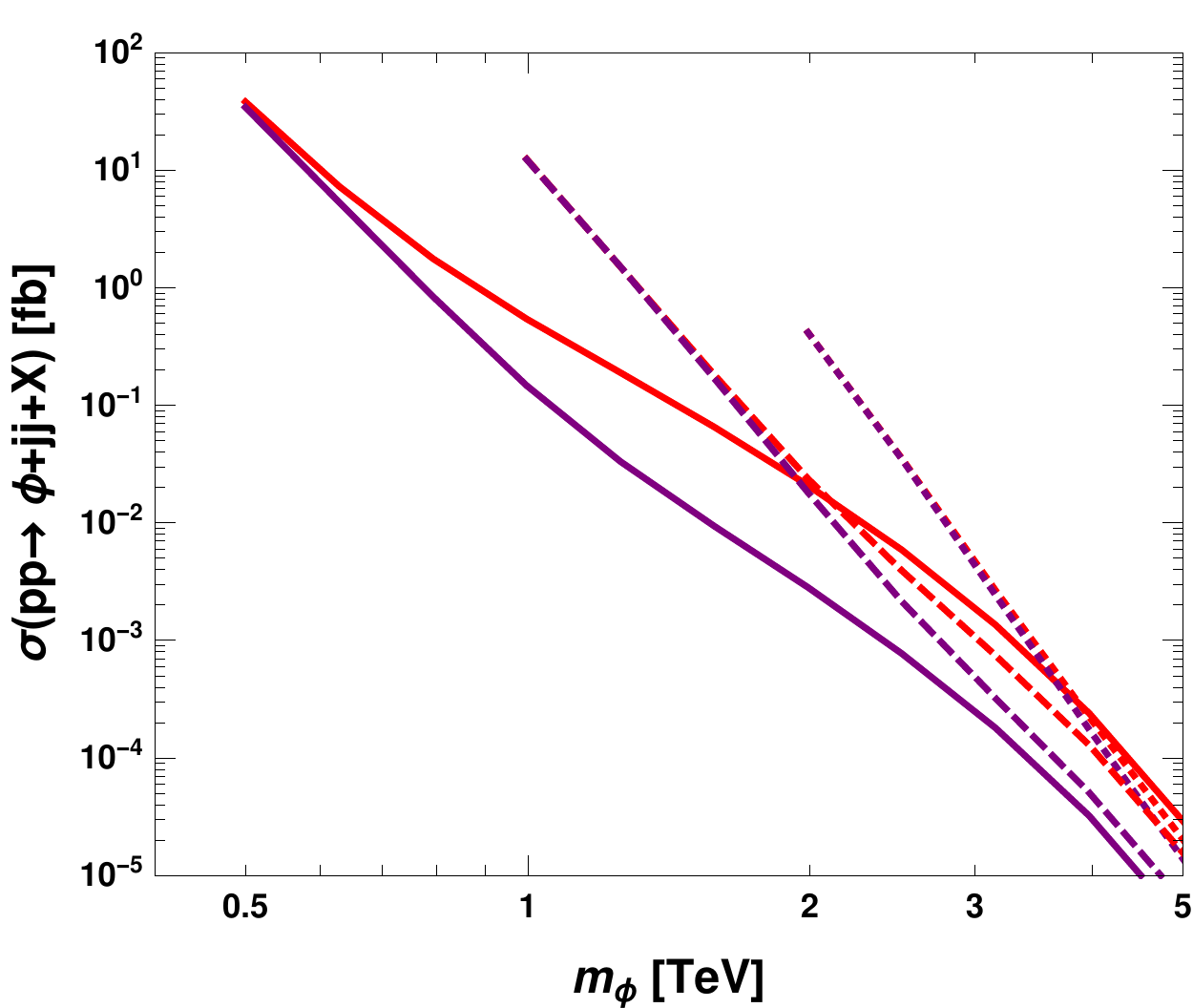}
\caption{ Global Higgs production rates via gluon fusion (left) and
vector-boson fusion (right), as a function of the global Higgs mass.
Red and purple lines correspond to $M_\psi=m_\phi$ and
$M_\psi=2m_\phi$, respectively.  Plain, dashed and dotted lines
correspond to $\lambda=0.2$, $\lambda=1$, $\lambda=3$, respectively.
\label{fig:XS}}
\end{figure}

In order to asses the interplay of these production modes, we simulate
the production rates using {\tt MadGraph5} \cite{MG}, based on a {\tt
FeynRules} \cite{FR} implementation of the global Higgs Lagrangian.
The parton density function set used is NN23LO1 \cite{Ball:2013hta},
with a factorization scale set to $\mu_F=m_\phi$.  The ggF and VBF
production rates are shown in Fig.~\ref{fig:XS} in the cases
$M_\psi=m_\phi$ (red curves) and $M_\psi=2m_\phi$ (purple curves).
All the bounds on the parameters described in Sec.~\ref{se:coup} are
taken into account.  In particular, for a given $\lambda$, the global
Higgs mass is bounded from below by $m_\phi>\sqrt{2\lambda} f$ , where
$f=800$~GeV.

In the VBF case, the dominance of the loop induced operators $\phi
(V^{\mu\nu})^2$ over $\phi |D^\mu H|^2$ can be recognized by the
cross-section dependence with respect to the heavy fermion mass
$M_\psi$.  This feature tends to happens for small $\lambda$, as
expected.  Also, the VBF rate is much smaller than the ggF rate at
small $\lambda$, while it dominates at large $\lambda$.  The crossover
occurs around $\lambda\sim 1$.

We see that the total production rate is high enough to motivate a
more precise study of the LHC implications of the presence of a global
Higgs.  In the following, since the VBF process is important only for
large $\lambda$, we choose to focus on a $ \phi + Y$ final state,
without requiring forward jet tagging.

\vspace{3mm} 
The LHC signals of the global Higgs can be split into two broad cases:
\begin{itemize}
\item Case I: All decays involving fermion resonances are closed.  The
phenomenology is then largely independent of the details of the heavy
fermion sector, and the narrow width approximation applies.  We study
this case in Sec.~\ref{se:diboson}.  

\item Case II: Some decays involving fermion resonances are open, and
the phenomenology depends strongly on the realization of the fermion
sector.  Some generic aspects of this case will be discussed in
Sec.~\ref{se:top_partners}.
\end{itemize} 

\section{Global Higgs Discovery Prospects: Decays into SM Particles} 
\label{se:diboson}

In this section we provide an estimate of the LHC sensitivity for
detecting the global Higgs at a center-of-mass energy of 13 TeV and
with 300 fb$^{-1}$ of integrated luminosity, assuming that all decays
involving fermion resonances are kinematically forbidden.  We will
take $M_\psi=m_\phi$ for definiteness, and we will therefore present
our results in the $m_\phi-\lambda$ plane.  The main decay channels to
be investigated, $\phi \to hh, ZZ, W^+W^-, t\bar{t}$, were discussed
in Fig.~\ref{fig:BRzones}, which shows the dominant channels in
different regions of parameter space.  Here we explore them in more
detail.

\subsection{The Hadronic NGB Channel}

We start by considering the case where the global Higgs decays
dominantly into NGBs, i.e.
\be
\phi\rightarrow W_LW_L, Z_LZ_L, hh\,.
\ee
The $W_LW_L$, $Z_LZ_L$ or $hh$ final states will decay further and
therefore there is a variety of final states that can be considered.
Decays into leptons could in principle provide very clean signatures.
Since the overall leptonic branching fractions are rather small we focus on fully hadronic decay modes, which may be more
relevant for discovery.\footnote{
In the context of resonant diboson searches, it has been noted that the fully hadronic channel has a slightly better sensitivity to high mass resonances than  other channels, see \textit{e.g.} Ref.~\cite{spaniards}. However it would be  certainly worth investigating other decay channels of the global Higgs.
Based on current experimental sensitivities,  promising final states  include $WW\rightarrow l\nu jj$, $ZZ\rightarrow 4l$ and $hh\rightarrow bb \gamma\gamma$.
}  
 The branching fractions of the $WW$, $ZZ$ and
$hh$ states into fully hadronic final states are all roughly $50
\%$.\footnote{The branching fraction of the $hh$ state into four
bottom quarks is roughly $30 \%$, but we choose not to consider the
possibility of $b$-tagging since the overall efficiency required for
four $b$-tags is around $1\%$ \cite{Aad:2015uka}.  }
Extrapolations of 8 TeV LHC bounds in the leptonic channels can be found in Ref.~\cite{Buttazzo:2015bka}. 

Before going into the details of the analysis it is worth pointing out that the $WW$ and $ZZ$ channels are also one of the main discovery channels for spin-1 resonances in composite Higgs models. Should a resonance be detected in this channel, a more detailed analysis will be required to discriminate between these cases. One such possibility is to look for specific channels that are forbidden in the spin-1 case, such as the decay into two photons (which is not allowed because of the Landau-Yang theorem), or the decay into two Higgses (which is forbidden because of Bose symmetry). Secondly, neutral spin-1 states typically come together with charged ones that are only split in mass by electroweak breaking effects, while possible partners of the Global Higgs are split by the larger $SO(5)\to SO(4)$ breaking.
Finally, the final-state angular distribution can be used to discriminate the spin of the decaying particle, which would of course be a rather challenging task. We will not discuss further these possibilities in this paper, but rather focus on the LHC phenomenology of the Global Higgs alone.

Since we are interested in the case where $m_\phi \gg m_W,m_Z,m_h$,
the produced $W$, $Z$ and $h$ are typically highly boosted and their
hadronic decay products are collimated in the detector frame, forming
a single, large-radius jet.  These are usually called fat jets in the
literature.  In order to maximize the signal rate, we suggest
searching for these fat jets.  In the following, a fat jet is denoted
by $J$ while a standard jet is denoted by $j$.  The process we are
interested in is thus \footnote{ We focus on the dominant gluon-fusion
production mechanism leading to a $JJ$ final state.  In the VBF mode,
one may expect the extra information from the two forward jets to be
useful for further background rejection.  However, to the best of our
knowledge the production of a resonance through the VBF mechanism
followed by decays into two fat jets, \ie~the $JJjj$ final state, has
not been investigated by the LHC collaborations.  }
\be 
pp  \rightarrow \phi^*  \rightarrow JJ ~.
\ee

The fully hadronic analysis is very challenging.  We describe below a
simple way to estimate the LHC reach for discovery of the global Higgs
in these modes.  We will rely on the recent progress accomplished with
jet substructure techniques \cite{Butterworth:2008iy} which show a
promising potential for QCD background rejection.  Such techniques
have been applied to the search of pairs of boosted weak bosons by the
ATLAS collaboration in the fully hadronic channel~\cite{ATLAS_note},
and we will use some of their results, especially the efficiency of
tagging boosted gauge bosons in the jet samples.

The signal is computed using our implementation of the effective
operators discussed in Sec.~\ref{se:coup}.  For a resonance decaying
into either $ZZ$ or $WW$ states, the signal efficiency for the
corresponding diboson-tagged hadronic final states has been estimated
in Ref.~\cite{ATLAS_note} at $9- 10 \%$ (with a 20\% uncertainty).
This efficiency includes the tagging as either a $[ZZ]$ selection or
as a $[WW]$ selection, as defined by ATLAS \cite{ATLAS_note}, while we
only require tagging as a diboson event, which we denote as $[VV]$.
Using the jet-tagging conditional probabilities computed in
Ref.~\cite{Fichet:2015yia} (see also \cite{Allanach:2015hba}) we can
substitute the $[WW]$ or $[ZZ]$ tagging for a $[VV]$ tagging by
multiplying the $WW$ efficiency by $P(\,[VV]\,|WW)/P(\,[WW]\,|WW)$ and
similarly for the $ZZ$ efficiency.\footnote{ In the conditional
probability $P(X|I)$, $I$ denotes the true event before tagging, and
$X$ labels the selection, that we write here between brackets.  For
our purposes we need
$P(\,[VV]\,|WW)=P(\,[V]\,|W)P(\,[V]\,|W)=0.65^2=0.43$,
$P(\,[WW]\,|WW)=(P(\,[W]\,|W) +P(\,[W/Z]\,|W))(P(\,[W]\,|W)
+P(\,[W/Z]\,|W))=0.38$, using the tagging probabilities of
Ref.~\cite{Fichet:2015yia}.  Similarly we find $P(\,[VV]\,|ZZ)=0.51$
and $P(\,[ZZ]\,|ZZ)=0.37$.  } The efficiencies obtained in this way
are $\sim 11-12 \%$, so that the overall efficiency that allows for
both $WW$ and $ZZ$ final states, without trying to tell them apart,
turns out to be similar to the efficiencies found in the ATLAS
analysis.  We assume that these efficiencies will not change
significantly in the 13 TeV run, and we use $12\%$ for the $WW$ and
$ZZ$ channels as well as for the $hh$ channel.  ATLAS also estimates
the average background selection efficiency of the tagger in simulated
QCD dijet events satisfying the same cuts to be roughly $0.01 \%$,
showing the power of the jet substructure tools.

We turn now to a more detailed discussion of the dominant QCD
background.  In order to obtain a realistic dijet background for this
search, the whole process of jet reconstruction, grooming, filtering
and tagging should be accurately simulated.  As an alternative to a
complete simulation, we estimate the $JJ$ background at 13~TeV from
the $JJ$ background obtained in the 8 TeV dijet analysis by
ATLAS~\cite{ATLAS_note}.  We describe next how to obtain both the
shape and the normalization of the 13 TeV dijet background.

Let us start with the background distribution shape, expressed as a
function of the invariant mass of the reconstructed dijet system
$m_{JJ}$.  The observed distribution was fit by
ATLAS~\cite{ATLAS_note} to an analytic function $f(m_{JJ}/\sqrt{s})$,
so that the $m_{JJ}$ distribution scales roughly as the center-of-mass
energy.  We have checked this scaling behavior with a parton-level
simulation of dijet production.  Hence, we can use the background
shape of the 8 TeV analysis with a simple rescaling
$f(m_{JJ})\rightarrow f(13/8 \,m_{JJ})$.  One should note that the
ATLAS analysis of the 8 TeV data involves a $p_T$ cut on the leading
jet, $p_T(j)>540$~GeV. This cut leads to $m_{JJ}>1080$ GeV since a cut
$p_T > p_T^{\rm min}$ implies $m_{JJ} > 2 p_T^{\rm min}$, and also
slightly deforms the $m_{JJ}$ distribution at low invariant mass.
Therefore, in order to extrapolate the background from 8 to 13 TeV, we
also have to rescale the $p_T$ cut on the leading jet by $13/8$, thus
taking $p_T>877$ GeV. This in turn implies a cut $m_{JJ}>1754$ GeV at
$\sqrt{s}=13$~TeV.

With the shape determined as above, we have to fix the overall
normalization of the dijet background at 13~TeV. We need first the
total number of events obtained after tagging two jets as weak bosons
in the ATLAS 8 TeV analysis.  The total number of events has been
reported in \cite{ATLAS_note} in three overlapping categories: $WW$,
$ZZ$, $WZ$, with $\hat n_{WW}=425$, $\hat n_{ZZ}=333$, $\hat
n_{WZ}=604$.  The statistics of the overlapping event numbers for the
$n_{WW}$, $n_{ZZ}$, $n_{WZ}$ categories has been thoroughly studied
in~\cite{Fichet:2015yia}.\footnote{These variables follow a joint
trivariate Poisson distribution.} Knowing the tagging probabilities, a
simple likelihood analysis like the one described
in~\cite{Fichet:2015yia} provides the underlying number of jets before
tagging, ${\hat n}_{JJ}=107539$.  This number can then be multiplied
by the total mis-tagging probability of QCD jets into weak bosons
$P([VV]|JJ)=P([V]|J)^2=6.4\times 10^{-3}$ obtained
in~\cite{Fichet:2015yia}, giving the overall normalization of the 8
TeV dijet background for hadronically decaying dibosons: ${\hat
n}_{JJ}(8\, {\rm TeV})= 688$.  This allows us to estimate
$\sigma_{JJ}^{\rm ATLAS}(8\, {\rm TeV})=33.9$~fb.

We stress that this number corresponds to events with a $p_T(j)>540$
GeV cut.  In order to proceed with the extrapolation, we rescale this
number by the ratio of partonic cross-sections from 8 and 13 TeV,
including the rescaled $p_T$ cut discussed above,
\be
\sigma^{\rm parton}_{JJ}(13 \,{\rm TeV})/\sigma^{\rm parton}_{JJ}(8\, {\rm TeV})~\approx~0.3~.
\ee 
One notices the well-known feature that this ratio is smaller than one
-- see e.g. the general LHC cross-section
plots~\cite{Stirling_XS_plots}.  The total event rate at 13 TeV
extrapolated from the ATLAS analysis is then given by
 \be
 \sigma_{JJ}^{\rm ATLAS}(13\, {\rm TeV})~=~\sigma_{JJ}^{\rm ATLAS}(8\, {\rm TeV}) \, \frac{\sigma_{JJ}^{\rm parton}(13\, {\rm TeV})}{\sigma_{JJ}^{\rm parton}(8\, {\rm TeV})}~\approx~10 \; \mbox{fb}~.
 \ee
With this information, we have fixed the inferred $m_{JJ}$
distribution at 13~TeV with a cut $m_{JJ}>1754$~GeV. We will then
simply use the analytic fit to extrapolate the background to the
$m_{JJ}<1754$~GeV region.

\begin{figure}
\centering
\includegraphics[scale=0.8,clip=true, trim= 0cm 0cm 0cm 0cm]{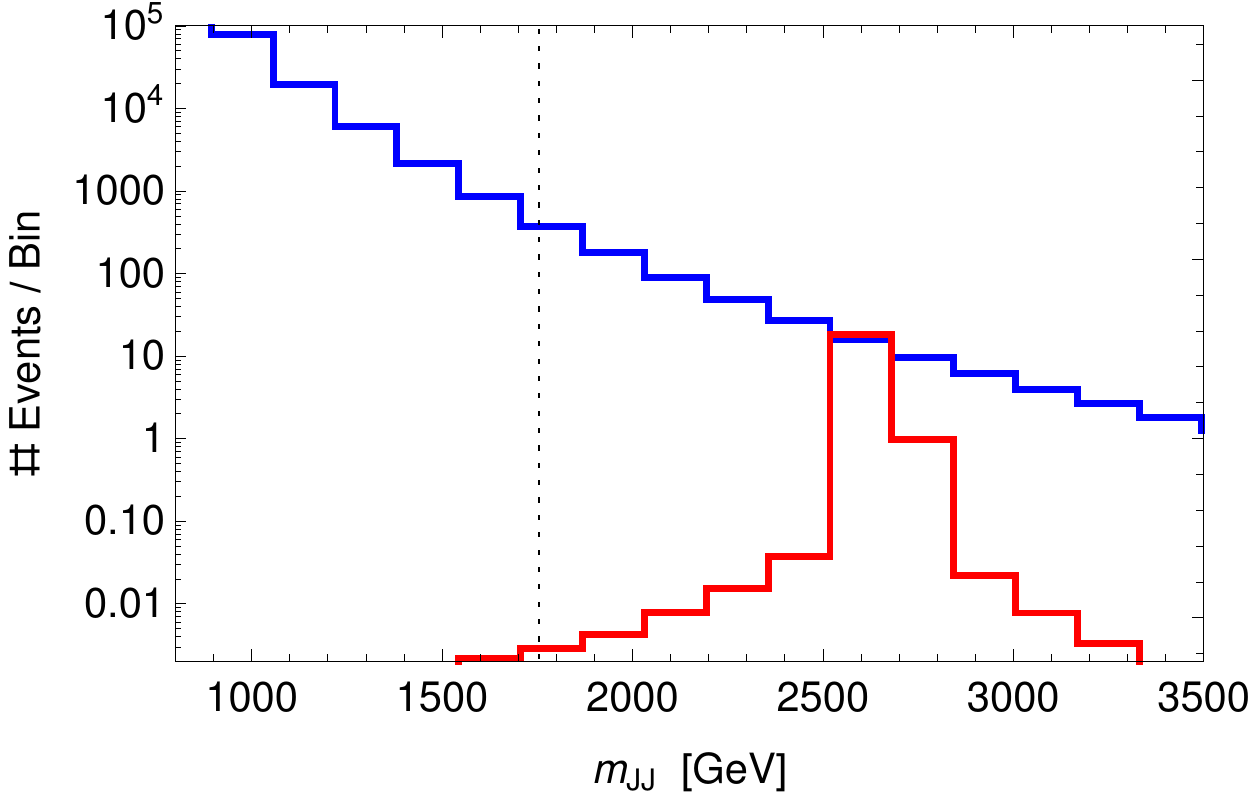}
\caption{ Projection of the dijet background at $13$ TeV extrapolated
from an ATLAS $8$ TeV analysis \cite{ATLAS_note}.  To the left of the
vertical dotted line the background is extrapolated using the model
obtained in that reference.  A signal for $pp\rightarrow \phi
\rightarrow JJ$ assuming $f=800$~GeV, $\lambda=0.2$,
$m_\phi=2640$~GeV is shown in red.
\label{fig:bkd}}
\end{figure}

Finally, various realistic improvements on background rejection based
on jet substructure techniques have been pointed out in
Ref.~\cite{Goncalves:2015yua}.  A simple improvement is to reduce the
radius of the cone algorithm for the first step of jet identification.
Indeed, the radius of a jet from weak bosons is typically $\Delta R
\sim m_V/p_T \sim 0.4$ at $8$~TeV. Using the simulation
of~\cite{Goncalves:2015yua}, we find that the mis-tagging rate
$P(V|J)$ can be reduced by a factor $\sim 0.5$, when taking $\Delta R
=0.4 $ instead of $\Delta R =1.2 $.  We will assume that this
improvement takes place, so that the dijet background is reduced by
$(0.5)^2$.  We regard our estimated background as roughly
representative of what will be obtained at the 13 TeV LHC run.  The 13
TeV extrapolated background can be seen in Fig.~\ref{fig:bkd}.

In order to assess discovery, we use an actual hypothesis test instead
of a p-value significance test.\footnote{The p-value criteria,
although widely used in particle physics, is also well-known for not
being a hypothesis test and can lead to erroneous results, see
Refs.~\cite{berger1,berger2}.} The background-only hypothesis is
denoted by $H_0$.  The hypothesis that a signal exists is denoted by
$H_1$ and is parameterized via $(m_\phi,\lambda)$.  The hypothesis
test we employ is the discovery Bayes factor
\be
B_0=\frac{P({\rm data}| H_1 )}{P({\rm data}| H_0 )}=\frac{\int  L(m_\phi,\lambda)\pi(m_\phi)\pi(\lambda)\,dm_\phi d\lambda }{L_\textrm{bg-only}}~,
\ee
where the likelihood function $L$ is obtained from the product of the
Poisson likelihoods in each bin, and we use flat logarithmic prior
density functions, $\pi$'s, for the $\lambda$ and $m_\phi$ parameters,
with ranges $\lambda \in [0.2,3]$ and $m_\phi \in [0.4,4]$~TeV,
respectively. The denominator $L_\textrm{bg-only}$ can be obtained from $L(m_\phi,\lambda)$ by taking $m_\phi \to \infty$.

Following Ref.~\cite{Cowan:2010js}, we assume that our projected data
have no statistical fluctuations (\ie~they are ``Asimov'' data)
arising from a signal with underlying parameters $(m_\phi',\lambda')$.
For each value of the parameters $(m_\phi',\lambda')$, one performs a
Bayesian discovery test to evaluate whether the signal contained in
these hypothetical data could be detected.  The discovery Bayes factor
applied to the projected data takes the form
\be
B_0(m_\phi',\lambda')=\frac{P({\rm data}(m_\phi',\lambda')| H_1 )}{P({\rm data}(m_\phi',\lambda')| H_0 )}~.
\ee

\begin{figure}
\centering
\begin{picture}(220,240)
\put(0,0){
\put(0,0){\includegraphics[scale=0.7,clip=true, trim= 0cm 0cm 0cm 0cm]{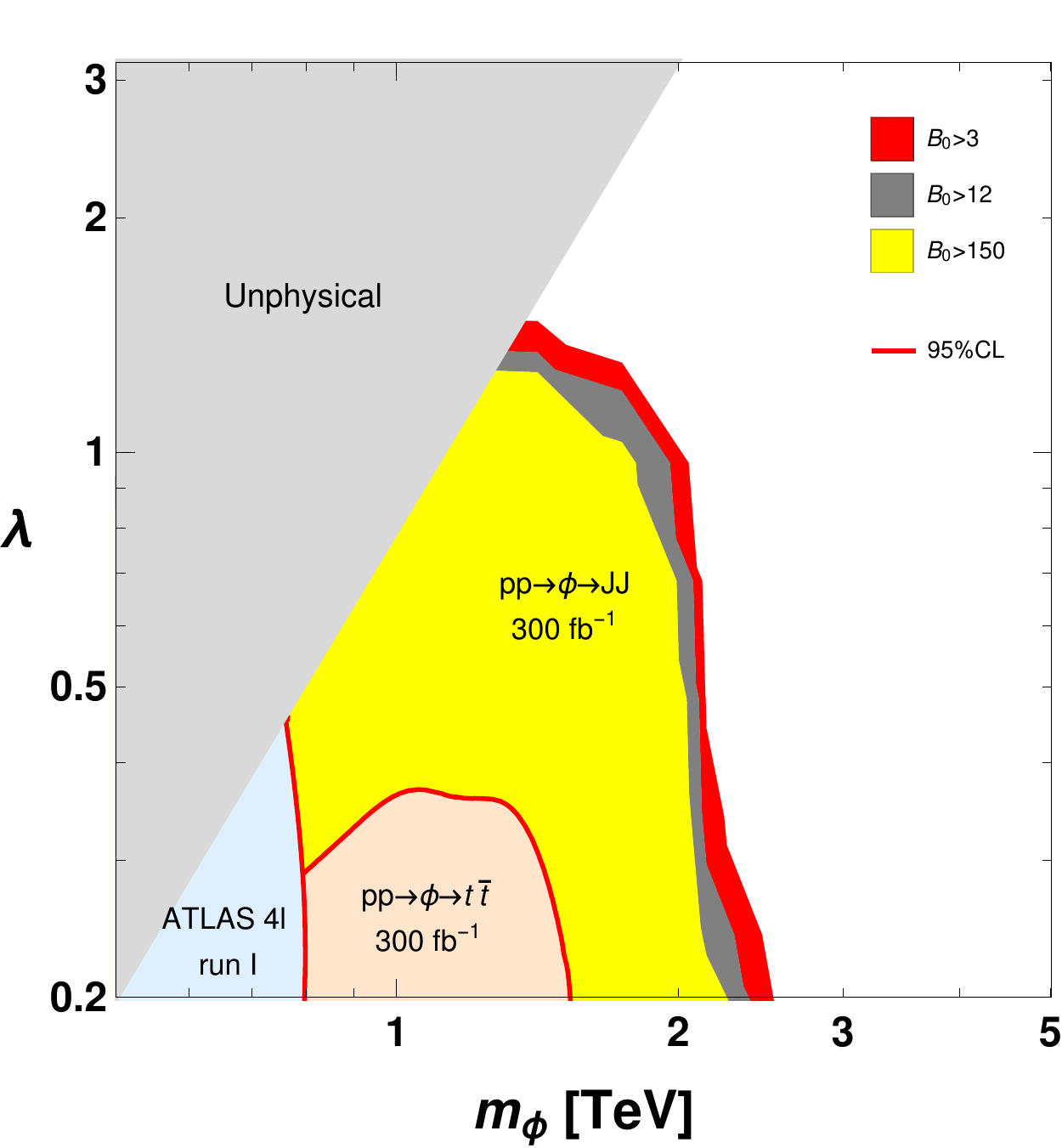}}
\put(55,190){\rotatebox{0}{ $ \Scale[0.7]{({\hat f}<f)}$}}
}
\end{picture}
\caption{Projected LHC sensitivities to a global Higgs signal with 300
fb$^{-1}$ at 13 TeV. The light blue region is a bound adapted from an
ATLAS heavy Higgs search \cite{ATLAS-CONF-2013-013}.  The light red
region is a projected 95$\%$CL limit from boosted top quark searches,
as extrapolated from Ref.~\cite{ATLAS-CONF-2016-014}.  The red, gray,
yellow regions show the discovery Bayes factor for the global Higgs in
the $pp\rightarrow \phi \rightarrow JJ$ channel, and correspond
respectively to weak, moderate and strong evidence for the signal
hypothesis.
\label{fig:ggF_analysis}
}
\end{figure}

The discovery Bayes factor for the global Higgs at the 13 TeV LHC run
with a luminosity of 300~fb$^{-1}$ is shown in
Fig.~\ref{fig:ggF_analysis}.  The threshold values $3$, $12$, $150$
can be roughly translated as $2$, $3$ and $5$ $\sigma$ significance
levels, respectively.

\subsection{The Boosted $t \bar t$ Channel}

Apart from NGBs, the other main decay channel of the global Higgs is
into top quark pairs,
\be
\phi\rightarrow t{\bar t}~.
\ee
This decay channel leads to boosted tops at the LHC. A recent search
for such resonant production of boosted top quark pairs has been
carried out by ATLAS using $3.2$ fb$^{-1}$ of 13 TeV data
\cite{ATLAS-CONF-2016-014}.  For our purpose of presenting a projected
sensitivity at 300 fb$^{-1}$, we extrapolate the \textit{expected}
95$\%$ CL bound on $\sigma\times BR$ given in
Ref.~\cite{ATLAS-CONF-2016-014}, which is obtained via a bump search
in the distribution of the mass of the reconstructed $t\bar t$ system,
$m_{t \bar t}^{\rm reco}$.

The extrapolation is done as follows.  We first assume that the
background event number is large enough that the counting statistics
in the bins of the $m_{t \bar t}^{\rm reco}$ distribution is
approximately Gaussian.  When this hypothesis is true, it implies that
the median expected $95\%$ CL limit as well as the associated error
bands can be extrapolated by rescaling the limit by a $\sqrt{3.2/300}$
factor.  This provides the projected $95\%$ limit at $300$ fb$^{-1}$
shown in Fig.~\ref{fig:ggF_analysis}.  We see that the region defined
by this limit corresponds to values of $m_\phi$ between $\sim 0.8$ and
$1.5$~TeV. We checked that the background in $m_{t \bar t}^{\rm reco}$
is sizeable, \ie~that the event number in each bin is at least ${\cal
O}(10)$, over the $[0.8,1.5]$~TeV range. Hence, the initial hypothesis
of Gaussian statistics is validated, and the extrapolation is
consistent.

\subsection{Results}

The projected sensitivities are summarized in
Fig.~\ref{fig:ggF_analysis}.  In the $JJ$ channel, we find that the
sensitivity reaches $m_\phi\sim 2-2.5$ TeV with 300 fb$^{-1}$, depending
on $\lambda$.  The sensitivity is greater for smaller $\lambda$,
reflecting the larger gluon fusion production rate, as explained in
Sec.~\ref{se:production}.  We also see that the boosted $t \bar t$
channel is less sensitive, with a mass reach of $m_\phi\sim
1.5$~TeV for low $\lambda$.  At larger values of $\lambda$ the
sensitivity of this search disappears because the ${\rm BR}(\phi \to
t\bar t)$ becomes suppressed (see Fig.~\ref{fig:BRzones}).

We emphasize that these sensitivities constitute only rough estimates,
based on extrapolations of specific experimental analyses.  This work
should be viewed as a first step towards a more realistic analysis.
Still, it is rather encouraging that these results appear to be
competitive with projected searches for top partners (for example, in
the recent analysis of Ref.~\cite{Backovic:2015bca}, the mass reach
for top partners is found to be around 1~TeV 
assuming $100$~fb$^{-1}$).  Therefore, there is a concrete possibility
that the global Higgs can be the first manifestation of compositeness
detectable at the LHC.

\section{Top Partners from Global Higgs Decays} 
\label{se:top_partners}

In this section we consider the case where the global Higgs can decay
into channels involving fermion resonances.  Of the large number of
resonances present in scenarios of the type described in
Sec.~\ref{se:coup}, one can reasonably expect that a subset of those
related to the top sector would be the lightest.  This is typically a
consequence of the large elementary-composite mixing characterizing
the top sector.  For definiteness, we will assume that only one of
those, which we call $t'$, is lighter than the global Higgs, so that
at most a few fermion channel are open:
\be
\phi\rightarrow t'\bar t ~(t\bar t') \,\qquad \phi\rightarrow t'\bar t'~.
\label{eq:phi_tt}
\ee 
Note that the branching fraction for the decays of
Eq.~\eqref{eq:phi_tt} can then be of order one, although most of our
analysis in this section is independent of this assumption.

In the following, we will allow the $t'$ state to be significantly
lighter than the global Higgs.  In this case, $t'$ will give a small
contribution to the loop-induced processes, in particular to the gluon
fusion process (as happens for the bottom quark contribution to the
Higgs-gluon-gluon coupling in the SM).  However, since it is only one
out of many states, our estimates for production studied in
Sec.~\ref{se:production} can be expected to remain roughly valid.

If several fermion resonances are significantly lighter than the
global Higgs, the latter is expected to become a rather broad
resonance, as pointed out earlier, with model-dependent branching
fractions.  Also, the $\phi gg$ coupling may be suppressed due to the
small loop functions.  Its size can also be rather model-dependent,
unlike the situation studied in Sec.~\ref{se:coup}.  For these
reasons, we do not consider such scenarios any further.

The $t'$ can have the following decays:
\be
t'\rightarrow  th,\ tZ,\ bW^+,
\label{tpdecays}
\ee again with highly model-dependent branching fractions
\cite{Backovic:2014uma}.  We will therefore focus on discussing the
broad features of searches for global Higgs decaying into $t'$, and
their interplay with standard $t'$ searches.

Depending on the experimental situation, the observation of the
channels described by Eqs.~(\ref{eq:phi_tt}) and (\ref{tpdecays})
would have slightly different consequences.  One can imagine, for
example, a scenario where the $t'$ state has already been observed at
the LHC, say through single production (or pair production by QCD, if
$t'$ is not too heavy).  Such vector-like quarks are expected in many
extensions of the SM, so that these particles alone cannot establish
unambiguously a composite Higgs scenario.  In that context, the
observation of the global Higgs would provide additional evidence in
support of the composite Higgs paradigm.  On the other hand, if $t'$
is heavy enough and the production rate of $\phi$ is sizeable, it may
be possible that the $t'$ themselves are easier to detect in the
global Higgs channel [i.e.~Eq.~\eqref{eq:phi_tt}] than in the standard
$t'$ production channels.  In addition, if the global Higgs decays to
$t'$ are the leading ones, which is plausible, the global Higgs
channel could even constitute the discovery channel for physics beyond
the SM. In either of these cases, the decay of the global Higgs into
$t'$s would have interesting consequences.

\begin{figure}
\centering
\includegraphics[scale=0.4,clip=true, trim= 3cm 3cm 3cm 3cm]{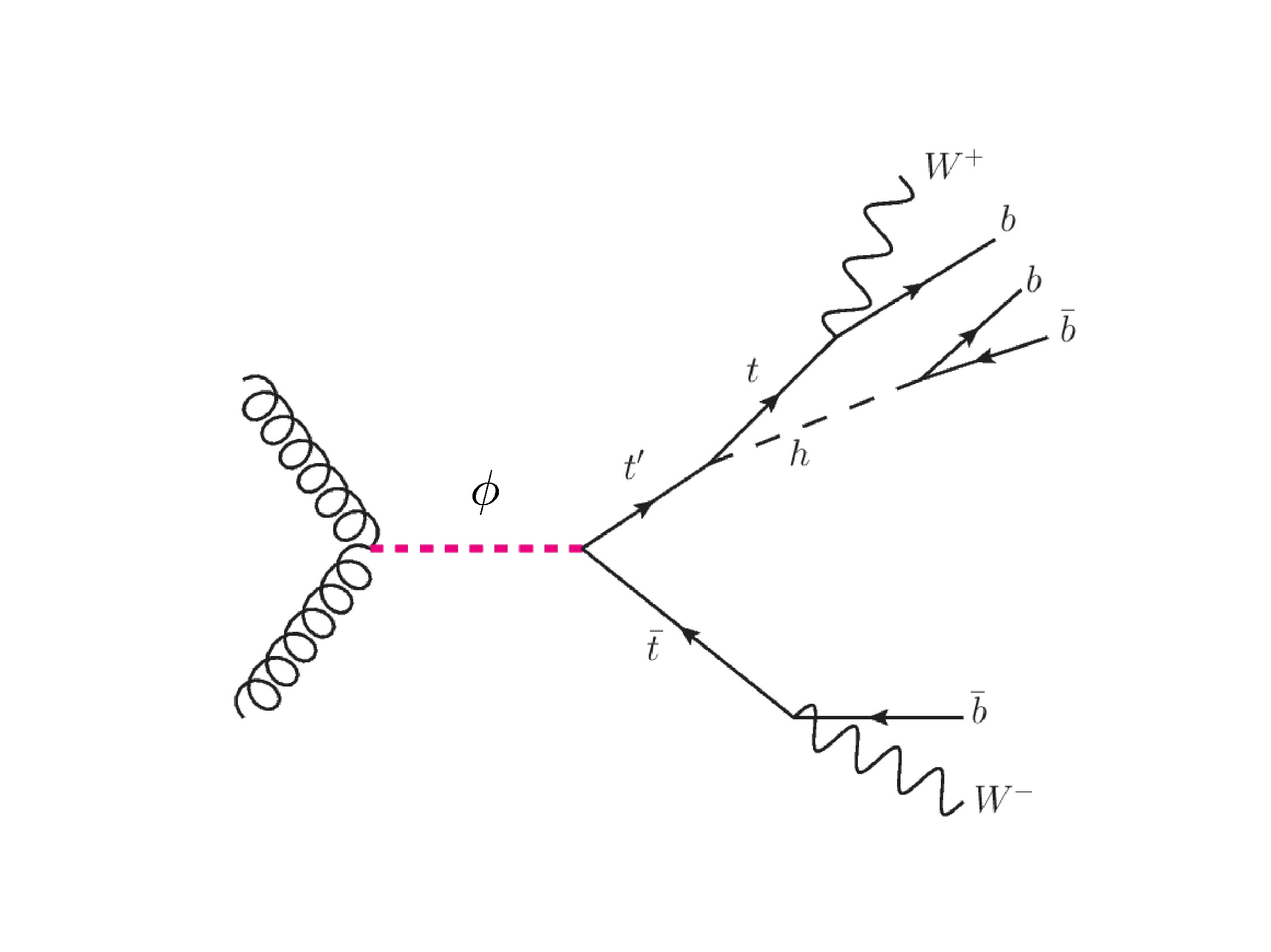}
\caption{ An example of global Higgs produced by gluon fusion and
decaying into a top quark and a top partner.  The $t'$ further decays
hadronically, hence the final states are potentially merged.
\label{fig:boosted_tp}}
\end{figure} 

As is well-known, light enough top partners can be pair-produced via
QCD processes, or produced singly, by the fusion of a $W$ and a $b$
quark, in association with a jet and a $b$-jet
\cite{DeSimone:2012fs,Azatov:2013hya,Backovic:2014uma,Matsedonskyi:2015dns,Backovic:2015bca}.
The former process is model-independent while the latter depends on
the strength of the coupling $ g /\sqrt{2}\, s_L\, \bar t_L' W^\mu
\gamma_\mu b_L$, where the mixing angle $s_L$ vanishes in the absence
of EWSB. At the $13$ TeV LHC, single $t'$ production is typically
expected to dominate over pair production when $m_{t'}$ is about a TeV
or above.  For reference, the production cross-section for a single
$t'$ of $1$ TeV is approximately
\be
\sigma_{t'}\approx 4.3 \, s_L^2 \,{\rm pb} \label{eq:sigma_tp}~,
\ee
at the 14 TeV LHC (using the results in
\cite{Backovic:2015bca}~\footnote{We thank the authors of
\cite{Backovic:2015bca} for clarifications regarding the cross section
Eq.~\eqref{eq:sigma_tp}.  }).  On the other hand, the $t'$ mass is
constrained by pair production searches at run I
\cite{ATLAS:2014pga,Aad:2014efa,Aad:2015kqa,Aad:2015gdg,Chatrchyan:2013uxa,
Khachatryan:2015axa,CMS:2014rda,CMS:1900uua,CMS:2014dka}.  The $95 \%$
lower bound on $m_{t'}$ is about $750-900$ GeV depending on the BRs.
We shall assume the conservative bound
\be 
m_{t'}>750~{\rm GeV} ~.
\ee 

The $t'$ decays offer several detection channels.  The channels with
highest branching fraction are the hadronic ones, $t'\rightarrow
t_{\rm had} Z_{ \rm had},\ b W_{\rm had},\ t_{\rm had}h_{\rm had} $.
However, these suffer from huge multi-jet, $b\bar b+$jets, $t_{\rm
had}\rm t_{\rm had}+$jets backgrounds in existing searches focussed on
either QCD pair or single $t'$ production.  Rather refined strategies
are often needed to tame the background, involving customized bottom
and top tagging, large missing $E_T$ cuts, and forward jet tagging.
In Ref.~\cite{Backovic:2015bca}, for the case of single $t'$
production, the most promising detection channels from each decay mode
have been found to be $t_{\rm had}Z_{\rm inv}$, $b W_{\rm lep}$,
$t_{\rm had}h_{bb}$.  The $t_{\rm had}h_{bb}$ channel requires careful
tagging techniques, and the signal drops to $~5\%$ after cuts.  Given
the cross section of Eq.~\eqref{eq:sigma_tp}, the production rate
after cuts may be matched by production via the global Higgs channel
that we discuss next.
 
Compared to the standard $t'$ searches, the production of $t'\bar t'$
and $t'\bar t \, (t \bar t')$ via decays of the global Higgs presents
a number of distinctive features, potentially useful in efficiently
eliminating the backgrounds.  First, the production is resonant, which
is not the case for usual $t'$ production modes.  The $t'\bar t'$,
$t'\bar t \, (t \bar t')$ are expected to be produced essentially
back-to-back, which provides a constraint on the topology of the
event.  Resonant production further implies that a shape analysis
(\ie~a ``bump search'') of the reconstructed $m_{t't'}$ ($m_{t't}$)
invariant mass can be carried out.  Second, in the case of $t'\bar t$
$(t\bar t')$ the top is highly boosted typically with
$\mathbf{p}_T\sim m_\phi/2$, so that these events are selected with
high trigger efficiency at ATLAS and CMS.
Third, if the $t'$ is significantly lighter than the global Higgs, the
$t'$ can be highly boosted.  This is in sharp contrast with SM $t'$
production, where the $\mathbf{p}_T$ of the $t'$ is typically small,
so that the decay products $th,\ tZ,\ bW^+$ are well separated.  One
may notice that for a boosted $t'$, the missing-energy based search in
the $t Z_{\rm inv}$ channel proposed in~\cite{Backovic:2015bca} does
not work, since the missing-$E_T$ from the neutrinos is not resolved
anymore.  However the high boost also opens up the possibility that
the hadronic decay products of the $t'$ itself can merge.  The object
to search for then becomes a single large-radius (\ie~``fat")
$t'$-jet.  This possibility has, to the best of our knowledge, never
been discussed in the literature.  Such fat jets should be analyzed
using jet substructure techniques.  As a basic first step, a grooming
technique (filtering \cite{BDRS}, pruning \cite{Ellis:2009su} ,
trimming \cite{Krohn:2009th}) can be used to remove extra jets from
pileup, soft radiation and the underlying event.  The remaining hard
subjets can then be used to reconstruct the $t'$ 4-momentum.
Combining this information with that of the other $t'$ or $t$ gives
then access to the global Higgs mass itself.

\begin{figure}[!tp]
\centering
\includegraphics[scale=0.48,clip=true, trim= 0cm 0cm 0cm 0cm]{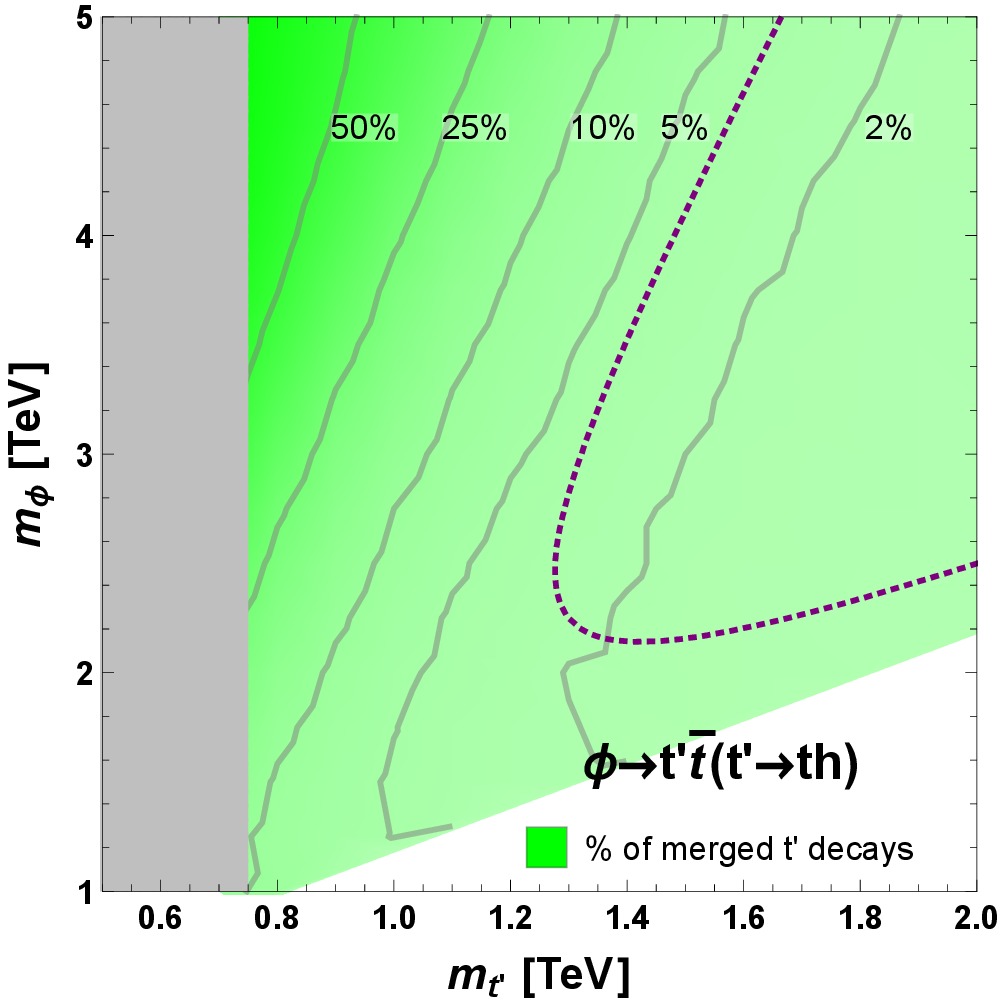}
\includegraphics[scale=0.48,clip=true, trim= 0cm 0cm 0cm 0cm]{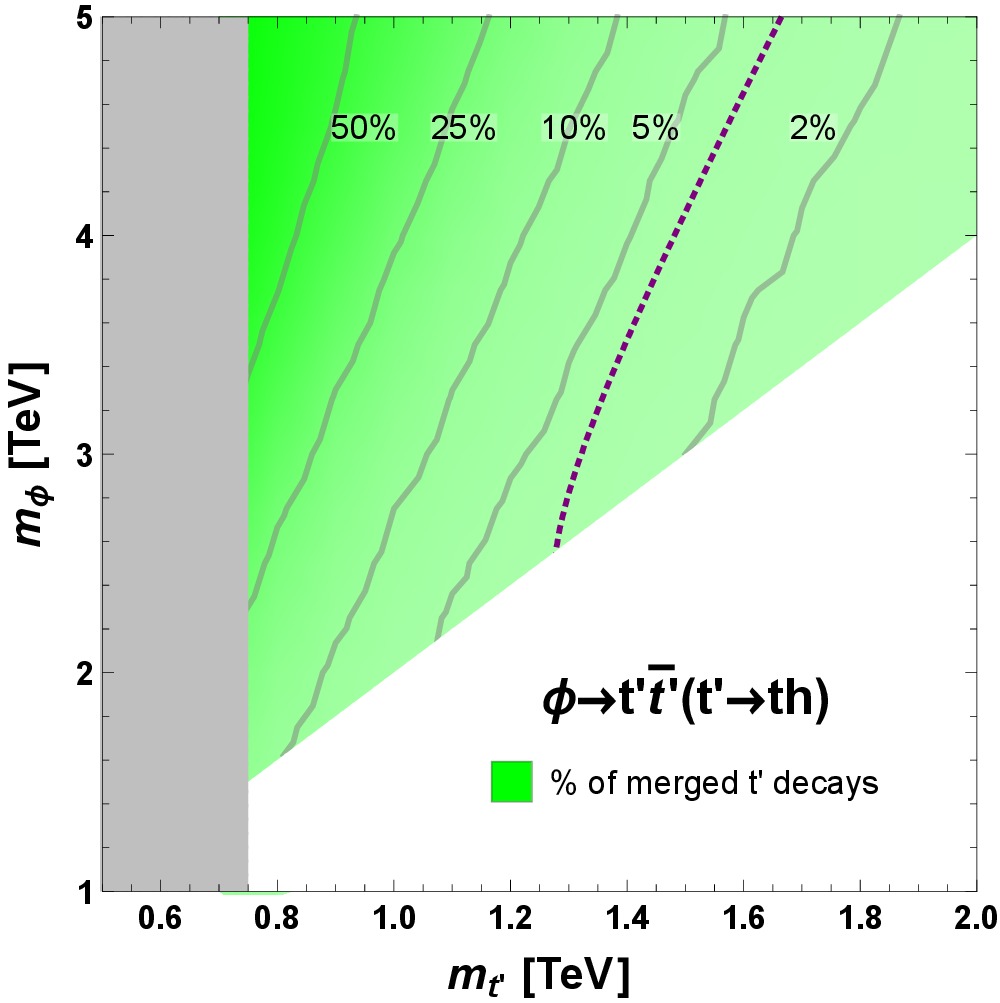} \\
\includegraphics[scale=0.48,clip=true, trim= 0cm 0cm 0cm 0cm]{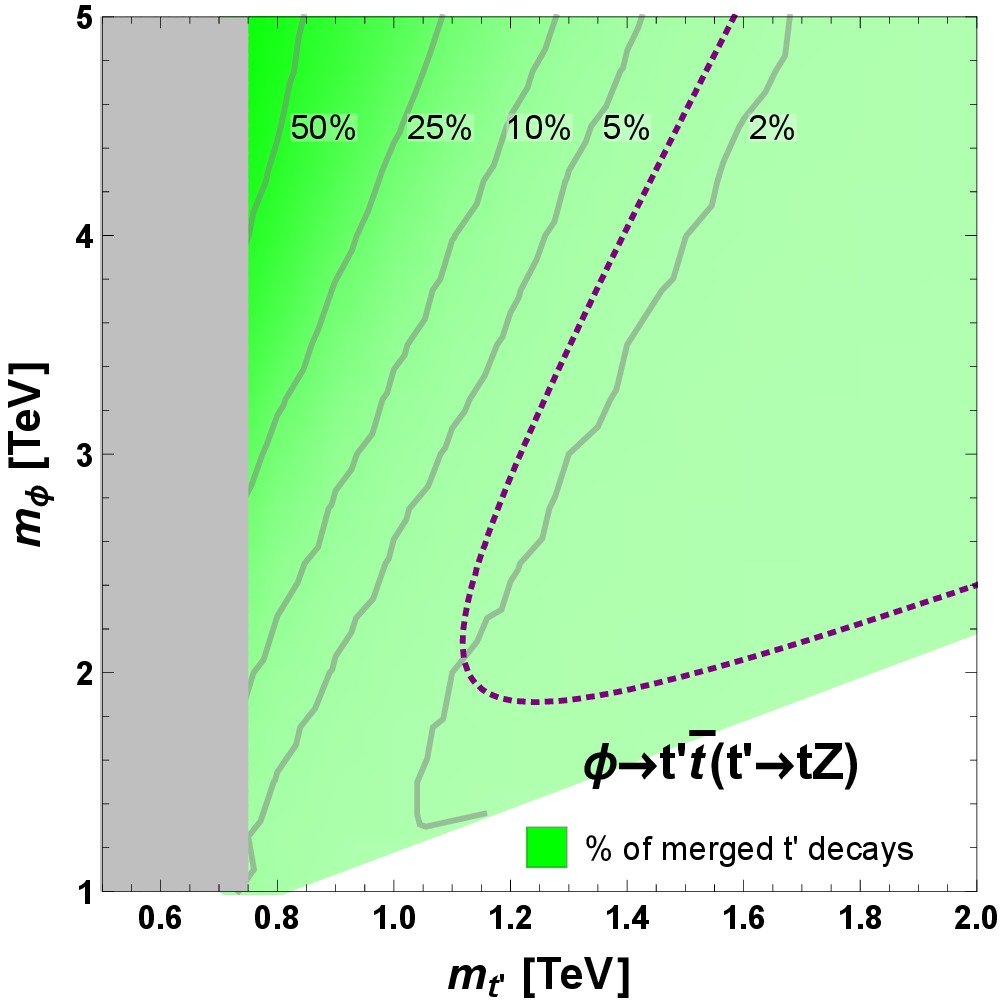}
\includegraphics[scale=0.48,clip=true, trim= 0cm 0cm 0cm 0cm]{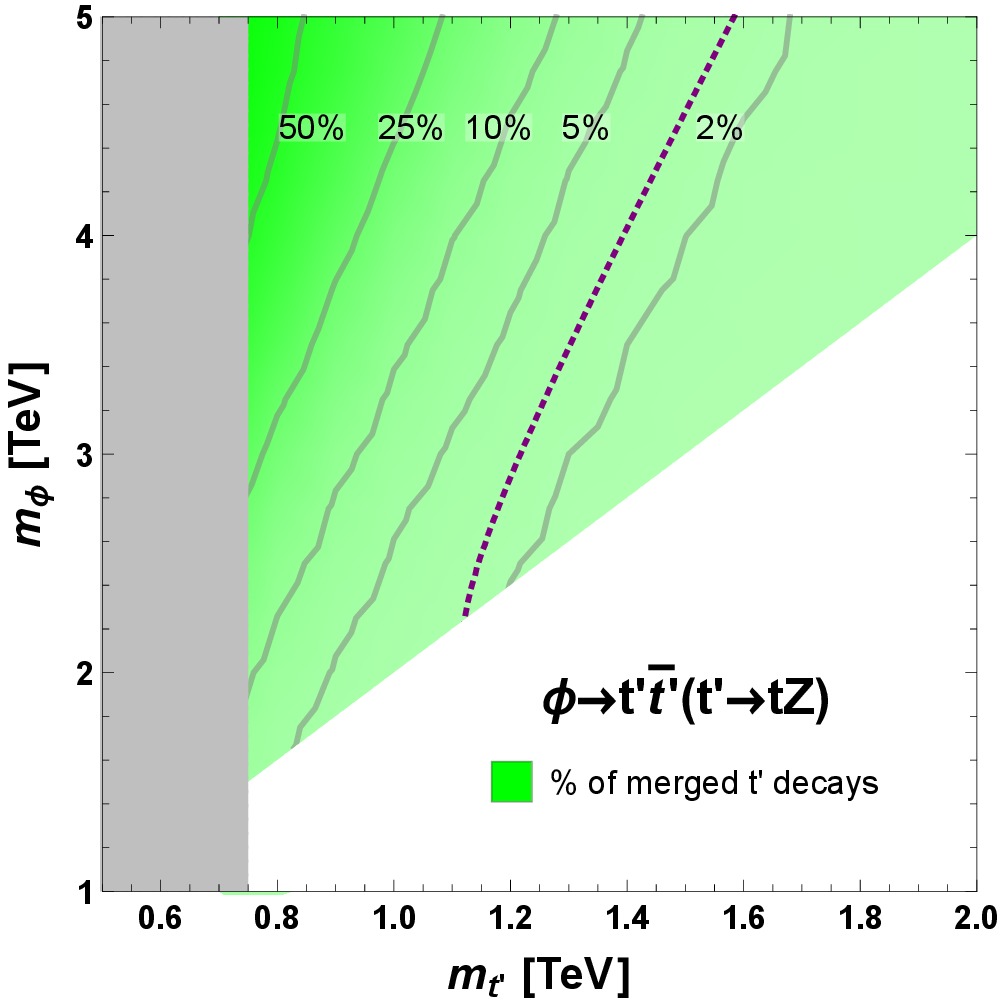} \\
\includegraphics[scale=0.48,clip=true, trim= 0cm 0cm 0cm 0cm]{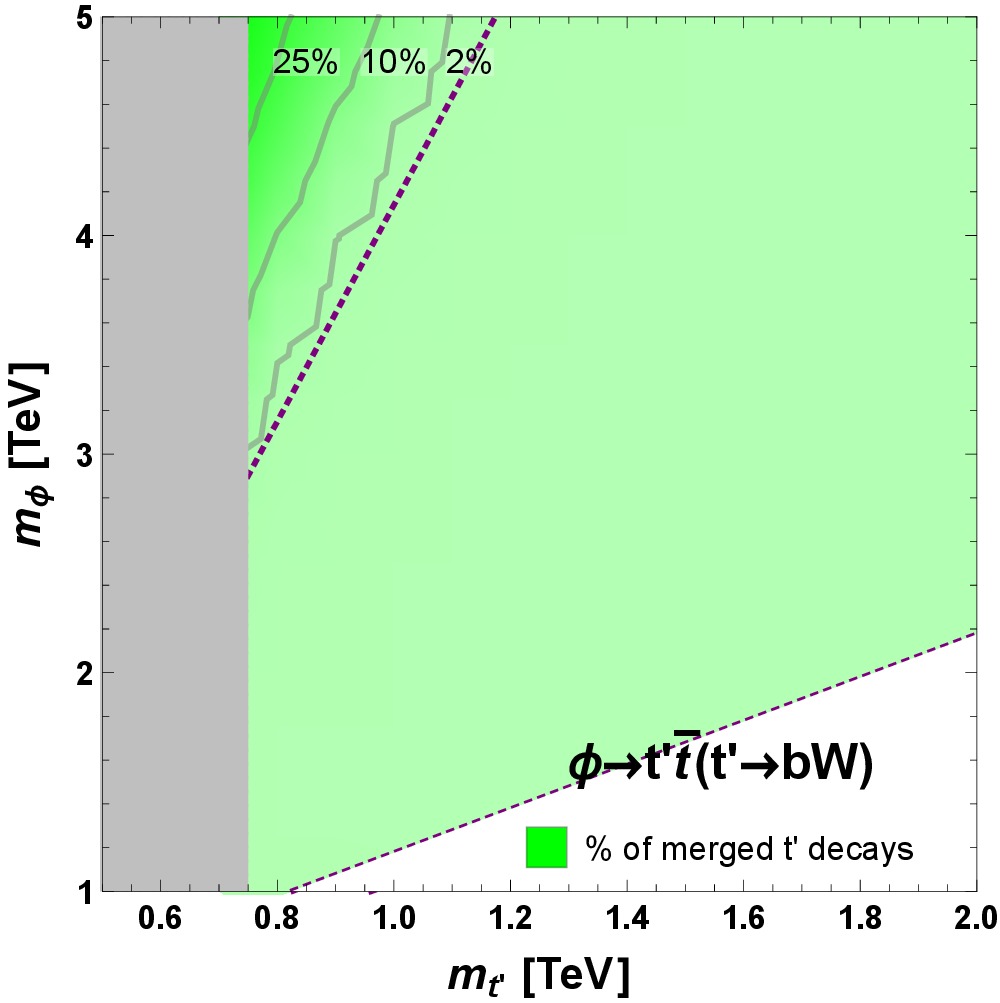}
\includegraphics[scale=0.48,clip=true, trim= 0cm 0cm 0cm 0cm]{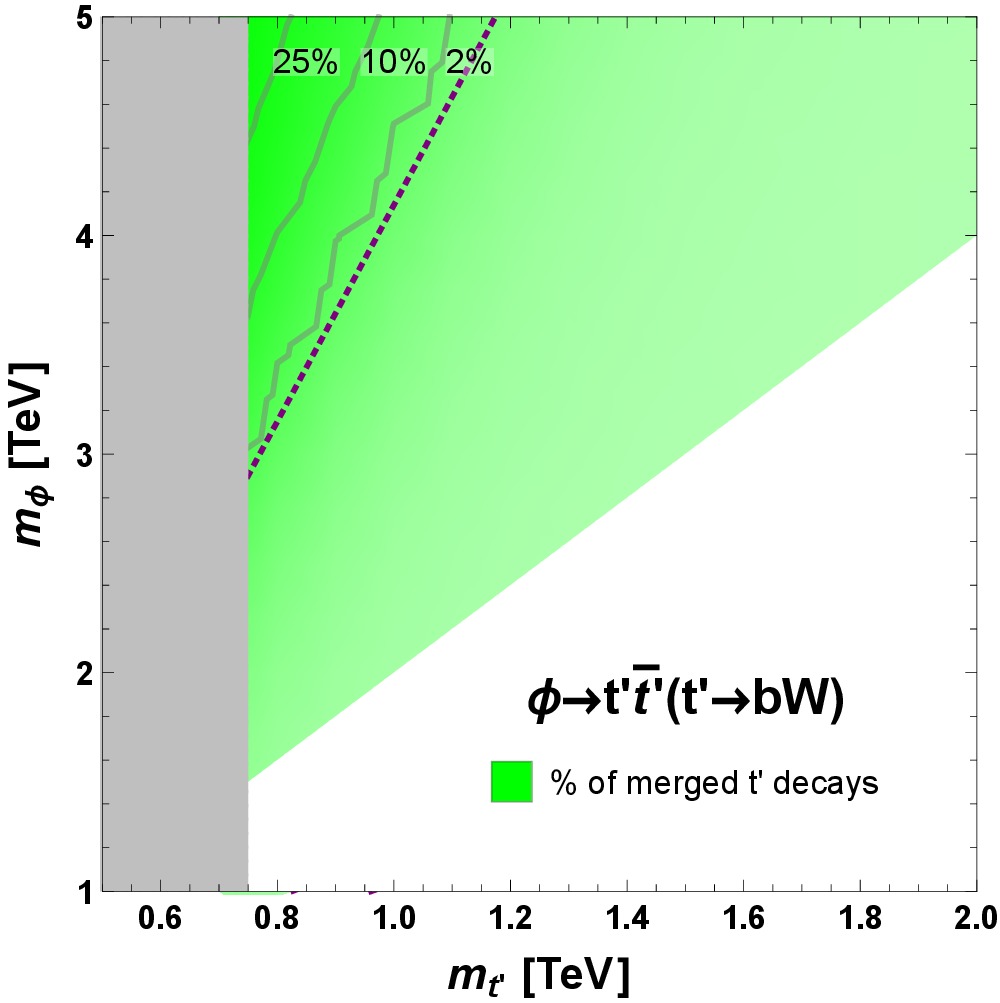} \\
\caption{Fraction of merged $t'$ decays in the $m_{t'}-m_\phi$ plane
for the cases of $\phi \to t' \bar{t}~(t \bar{t}')$ (left plots) and
$\phi \to t' \bar{t}'$ (right plots).  In the white region, these
decays cannot occur on-shell.  The plots from top to bottom correspond
to the possible $t'$ decays, $t'\to th$, $t'\to tZ$ and $t'\to bW$.
The gray vertical band is a conservative $95\%$ exclusion region from
Run I searches.  The dashed line is an estimate of the merging region
following the calculation of App.~\ref{se:analytic}, assuming
azimuthal $t'$ opening angle (see Eq.~\eqref{eq:DRphi}).
\label{fig:boost_TpTpbar}}
\end{figure}

Let us comment on the possible content of the $t'$-jet.  The merged
decay products from $b+ W$ resulting from a boosted $t'$ are similar
to a hadronic top decay with mass $m_t\rightarrow m_{t'}$.  The merged
$t+Z$ decays leads to a fat jet containing $b+2j+2j$, and the merged
$t+h$ contains to $b+2j+2b$.  These two last decay chains are more
likely to produce a fat jet, simply because there are more final
states that potentially overlap.  Besides, in the $t+h$ channel,
tagging the $b$ quarks inside the jet can dramatically reduce the
background.  This last channel is thus particularly attractive.  In
order to reduce further the $t'$-jet background, tagging techniques
can in principle be adapted or developed.  Tagging directly the whole
$t'$ decay seems difficult, since the $t'$ mass is a priori unknown
and the event has many subjets to combine.  A less ambitious approach
could be to tag the heavy $W$, $Z$, $h$ and top subjets inside the fat
jets.  This can be carried out using for example the pruning tagger of
Ref.~\cite{Ellis:2009me}.  Notice that the uncertainty on the
reconstructed subjet masses with this technique is about $\pm 10 $ GeV
\cite{Plehn:2011tg}, which implies that the $W$ and $Z$ cannot be
distinguished in such an approach.

A boosted $t'$-jet is an interesting object, both theoretically as it
may signal the existence of the global Higgs, and experimentally as it
leads to new channels to be analyzed with dedicated substructure
tools.  The remaining crucial question is ``How likely is it for
$t'$-jets to be produced from a global Higgs decay?''  To answer this,
we first notice that for a given production mode of the global Higgs,
the fraction of merged $t'$ decays depends only on the kinematics of
the global Higgs decay chain.  Therefore the fraction of merged $t'$
decays only depends on the global Higgs mass and the $t'$ mass, and
can be shown in the $m_{t'}-m_\phi$ plane irrespective of the details
of the model.

We evaluate the fraction of $t'$-jets by Monte Carlo (MC) integration.
We simulate the process of global Higgs production via ggF using {\tt
MadGraph5} \cite{MG} with our implementation of the global Higgs and
top partner Lagrangian in {\tt FeynRules} \cite{FR}.  We analyze the
six possible decay chains given by $\phi\rightarrow t'\bar t~(t
\bar{t'}) $, $\phi\rightarrow t'\bar{t'} $ followed by either
$t'\rightarrow t_{\rm had} Z_{ \rm had},\ b W_{\rm had}$, or $t_{\rm
had}h_{bb} $.  Denoting schematically $t'\rightarrow AB$, the fraction
is obtained by requiring that at least one of the jets from $A$ is
separated from a jet from $B$ by $\Delta R(A,B)<0.8$.  This is done
using {\tt MadAnalysis5} \cite{MA}.\footnote{At the LHC, the typical
radius of a QCD jet is $R\sim 0.4$.  The hadronic decays of heavy SM
particles start to merge for a $\mathbf{p}_T$ of a few hundred GeV.
For $h\rightarrow b\bar b$ for example, the threshold $\mathbf{p}_T$
is found to be $300\pm5$ GeV using the formulas of
App.~\ref{se:analytic} and asking for $\Delta R_{bb} <0.4+0.4 $ (see
Ref.~\cite{Negrini:2015qoa} and references therein).  } 
Using only  this condition on $\Delta R(A,B)$ leaves in principle the possibility of having resolved decay products within $A$ or $B$. When this happens, one obtains a ``partially-merged'' object, which is in principle also interesting. However we checked that in practice, depending on the process under consideration, the fraction of fully-merged events ranges among $\sim 90\% - 100\% $. In the following, we do not distinguish between these two subcases and refer to them simply as ``merged decays''.

\begin{figure}[t]
\centering
\includegraphics[scale=0.55,clip=true, trim= 0cm 0cm 0cm 0cm]{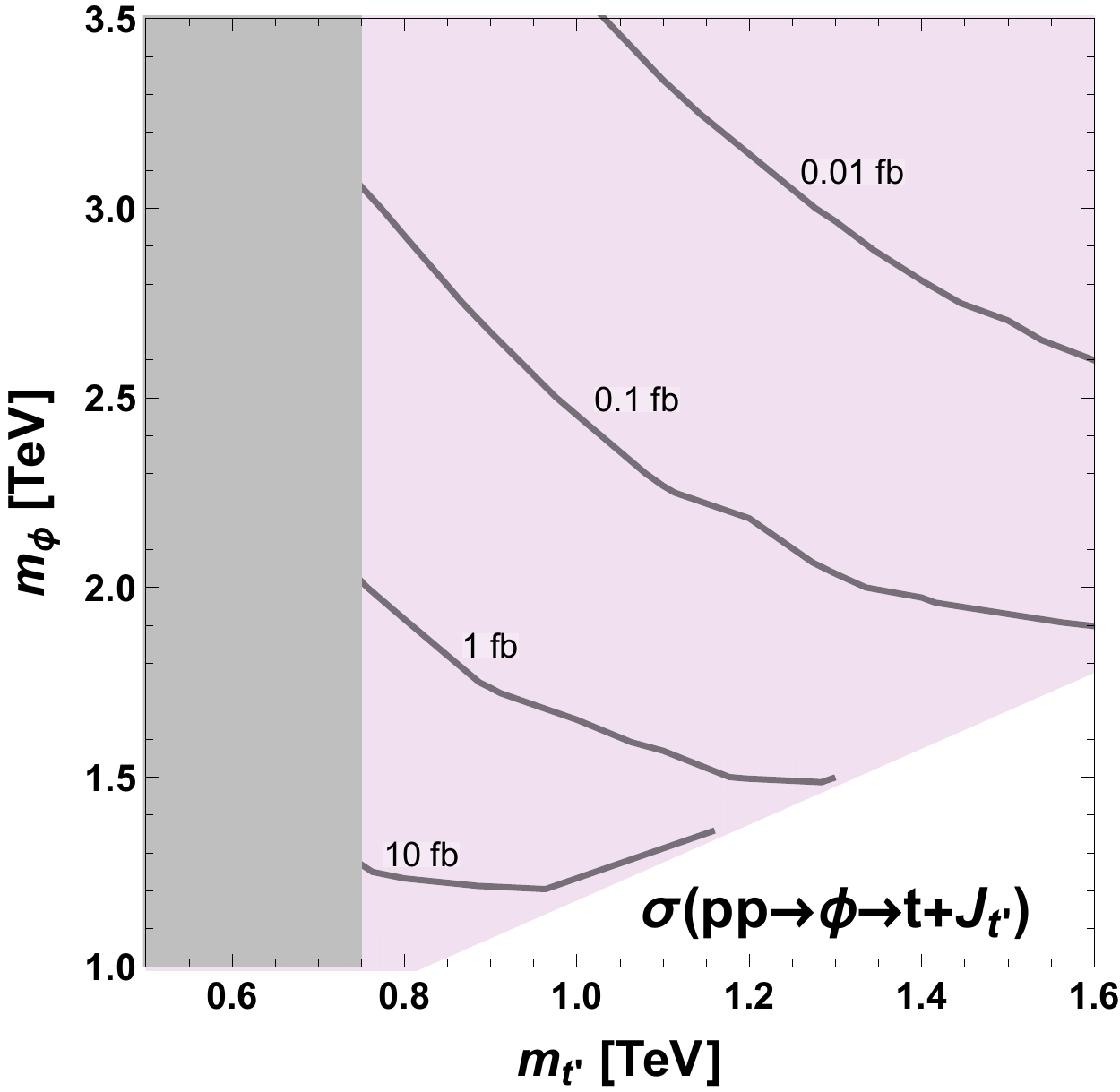}~~
\includegraphics[scale=0.55,clip=true, trim= 0cm 0cm 0cm 0cm]{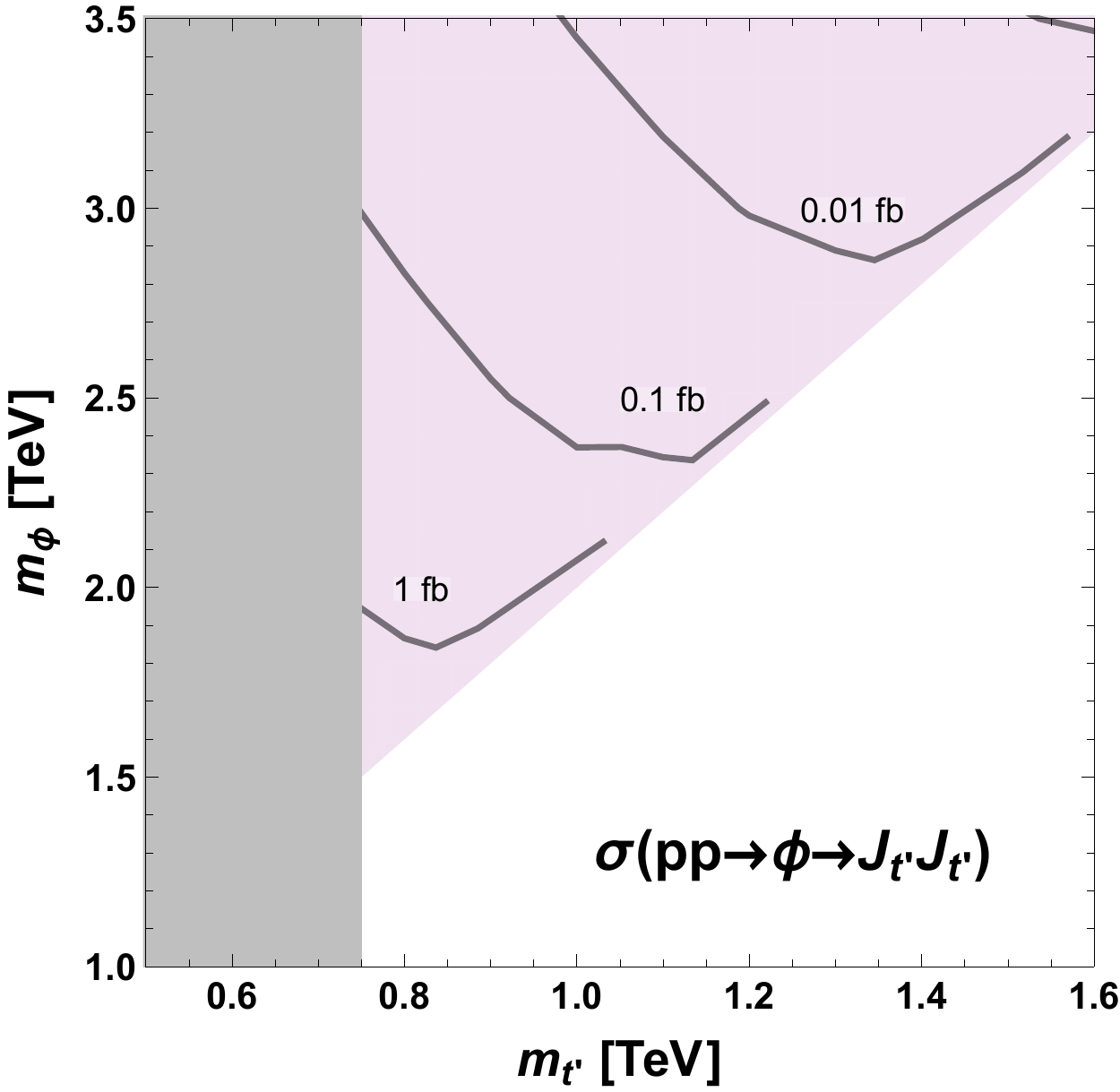}
\caption{ Single and pair production rate of $t'$-jets assuming
$2:1:1$ branching fractions for the $t'\rightarrow bW, tZ$ and $th$
channels.  We take $\lambda=0.2$ and assume $100\%$ decays of the
global Higgs into $t \bar t' (t' \bar t)$ (left plot) or into $t' \bar
t'$ (right plot).
\label{fig:XS_boosted_Tp}}
\end{figure}

The fraction of merged $t'$ decays in the $m_{t'}-m_\phi$ plane is
shown in Fig.~\ref{fig:boost_TpTpbar}.  We can see that in case of
$t'\rightarrow th$ and $tZ$ decays, a sizeable region features more
than $10\%$ of $t'$-jets.  On the other hand, in the case of $bW$
decay, the amount of $t'$-jets is smaller by an order of magnitude.
This is expected since the $b$ and $W$ jets have a smaller radius than
$t$, $Z$, or $h$ jets.  These features can also be understood
qualitatively using the analytic approach presented in
App.~\ref{se:analytic}.  The fraction of merged $t'$ decays obviously
increases with $m_\phi$ for a fixed $m_{t'}$.  However, the production
rate of the global Higgs drops with $m_\phi$.  In
Fig.~\ref{fig:XS_boosted_Tp} we show the expected cross section for
$t'$-jets, assuming the gluon fusion cross sections estimated in
Sec.~\ref{se:production}, and using the information of
Fig.~\ref{fig:boost_TpTpbar} with branching fractions for $t' \to bW$,
$t' \to tZ$, $t' \to th$ in the ratio $2:1:1$.  We see that the cross
sections are typically small.  Nevertheless, it can be interesting to
develop methods to detect these novel $t'$-jets.

\section{Conclusions } 
\label{se:conclusions}

In this paper we have performed an investigation of the LHC
signatures arising from the global Higgs, the ``radial" partner of the
NGBs identified as the SM Higgs and EW boson longitudinal
polarizations in modern composite Higgs constructions.

We evaluated the LHC sensitivity to global Higgs resonant production.
Our results suggest that these global Higgs channels can compete with
the standard searches for compositeness via SM production of top
partners.

In the case that the global Higgs decays mostly into NGBs and top
quarks, and not into fermion resonances, a projection at
$300$~fb$^{-1}$ of integrated luminosity for boosted hadronic channels
gives a sensitivity to the global Higgs up to a mass of $\sim
2.5-3$~TeV. We noted that this case is very predictive, effectively
depending on only two parameters $\hat f$, $r_v$.  Measuring both the
NGB and $t\bar t$ channels would provide an estimation of $r_v$.
Also, the $WW$, $ZZ$, $hh$ event rates are predicted to be in 2:1:1
proportions.

The case where the global Higgs can decay into fermion resonances is
much more model-dependent, hence we focused on a particular (but
well-motivated) scenario involving decays into charge $2/3$ top
partners.  The $t'$ produced through such resonant process may in
principle be easier to detect than the ones produced by standard SM
processes.  We also pointed out that in part of the parameter space,
such resonantly-produced $t'$ can be boosted enough to appear as a
single fat jet in the calorimeters.  We evaluated by MC simulation the
probability of having merged $t'$ decay products, and also provided an
analytic computation that approximately reproduces the boosted $t'$
regions.
 
Given these first encouraging results, it would be interesting to
further investigate the collider implications of the global Higgs.  In
particular, the rather striking possibility of getting boosted $t'$
states requires the development of new, dedicated jet substructure
analyses in order to properly select such signatures.

\acknowledgments We would like to thank Benjamin Fuks for
clarifications on the use of MC tools, and Thomas Flacke for useful
discussions.  This work was supported by the S\~ao Paulo Research
Foundation (FAPESP) under grants \#2011/11973 and \#2014/21477-2.
E.P. and R.R.~were partially funded by a CNPq research grant.

\appendix

\section{Loop Functions}
\label{loopfunctions}

For completeness, we collect here the well-known loop functions (see
\cite{Djouadi:2005gi}, for example) that appear at 1-loop order when
considering the couplings of a scalar to gauge bosons via heavy
fermion or spin-1 loops:
\bea
A_{1/2}(\tau) &=& 2 [\tau + (\tau -1) f(\tau)] \tau^{-2}~, \\ [0.3em]
A_{1}(\tau) &=& - [2 \tau^2 + 3 \tau + 3(2\tau - 1) f(\tau)] \tau^{-2}~,
\eea
where
\bea
f(\tau) &=& 
\left\{
\begin{array}{ll}
{\rm arcsin}^2 \sqrt{\tau}  &  \tau \leq 1   \\ [0.4em]
- \frac{1}{4} \left[ \log \frac{1 + \sqrt{1 - \tau^{-1}}}{1 - \sqrt{1 - \tau^{-1}}} - i \pi \right]^2 &  \tau > 1
\end{array}
\right.
~.
\label{floops}
\eea
In the limit that $\tau \rightarrow 0$, $A_{1/2}(\tau) \rightarrow
4/3$ and $A_{1}(\tau) \rightarrow -7$.

\section{An Analytic Estimation of the Boosted $t'$  Region} 
\label{se:analytic}

As a complement to the MC simulation above, we provide a purely
analytical technique to estimate the boosted $t'$ region.  Although
this approach is only qualitative as it provides only a region and not
a density, it has the advantage of being transparent and simple.
 
 We shall first set up some general kinematical expressions related to
 opening angles of decay products.  We consider a particle with mass
 $m$, arbitrary transverse momentum $p_T$ and rapidity $y$ decaying
 into two particles with transverse momentum $p_{T\,1}$, $p_{T\,2}$,
 and rapidities $y_1$, $y_2$, whose masses are neglected with respect
 to $m$ or $|p_T|$.  We are interested in the opening angle between
 the decay products, $(\Delta R)^2=(\Delta \eta)^2+(\Delta
 \phi)^2$.~\footnote{For massless particles the pseudorapidity $\eta$
 is equivalent to the rapidity $y$} An approximation that can be
 sometimes found in the literature is $\Delta R \approx 2 m / |p_T|$,
 which is only valid for $|p_T|\gg m$ and for symmetric decay
 configuration.  Here one needs to go beyond this case, so that we
 revisit the computation in order to establish well-controlled
 approximate formulas.

Using $p=q_1+q_2$ with transverse variables,\footnote{Namely
\be
\begin{pmatrix}
m_T \cosh y \\
\mathbf{p}_T \\
m_T \sinh y 
\end{pmatrix}=
\begin{pmatrix}
|\mathbf{p}_{T\,1}| \cosh y_1 \\
\mathbf{p}_{T\,1} \\
|\mathbf{p}_{T\,1}| \sinh y_1 
\end{pmatrix}
+
\begin{pmatrix}
|\mathbf{p}_{T\,2}| \cosh y_2 \\
\mathbf{p}_{T\,2} \\
|\mathbf{p}_{T\,2}| \sinh y_2 
\end{pmatrix}\,.
\ee
}
one obtains 
\be
m^2=2|\mathbf{p}_{T\,1}||\mathbf{p}_{T\,2}|(\cosh \Delta y - \cos \Delta \phi)\,, \label{eq:kin1}
\ee
where $\Delta \phi=\phi_2-\phi_1$ is the difference between the
azimuthal angles and $\Delta y =y_2-y_1$.  In order to go further
analytically, an extra condition needs to be chosen.  We find that two
different conditions independently lead to the same result.

A first condition is to select the particular configuration that gives
the minimal $\Delta R$ angle.  This lower bound is useful in order to
assess the radius for grooming algorithms, and will be needed in our
approach to jet merging.  Asking for the lowest $\Delta R$ amounts to
maximize the $|\mathbf{p_{T\,1}}||\mathbf{p_{T\,2}}|$ product.  Using
transverse momentum conservation, one obtains that
\be
  |\mathbf{p}_{T\,1}|=|\mathbf{p}_{T\,2}| = \frac{|\mathbf {p_T}|}{2 \cos(\Delta\phi/2)}\,. \label{eq:pT12}
\ee
Using this expression in Eq.~\eqref{eq:kin1} provides the main formula
\be
\frac{\cos^2( \Delta\phi/2)}{\cosh^2(\Delta y /2)}=\frac{ |\mathbf{p}_T|^2}{m^2+|\mathbf{p}_T|^2}\,.
\label{eq:kin2}
\ee
Alternatively, this equation can also be obtained starting from the
condition $ |\mathbf{p_{T\,1}}|= |\mathbf{p_{T\,2}}|$, which is
motivated by the fact that such symmetric configuration is
statistically the most likely to occur in the two-body decay.
Together with momentum conservation, the condition implies that
$y=(y_1+y_2)/2$ exactly, and Eq.~\eqref{eq:kin2} follows.  This
equation provides the kinematic bounds on $\Delta\phi$, $\Delta y$.
From \eqref{eq:pT12}, one can see that the minimal and maximal $
|\mathbf{p}_{T\,1,2}|$ are respectively equal to $|\mathbf{p}_{T}|/2$,
$\sqrt{m^2+|\mathbf{p}_{T}|^2}/2$, and correspond respectively to
$\Delta \phi =0$ and $\Delta y =0$.

The only assumption done at this stage is on the absolute value of
outgoing transverse momenta.  Assuming further that $\Delta \phi \ll
1$ and $\Delta y \ll 1$, Eq.~(\ref{eq:kin2}) implies that
$m\ll|\mathbf p_T|$ and it then follows that
\be
\Delta R = \frac{2 m}{{|\mathbf{p}_T|}}
+O\left(\Delta y^4, \Delta\phi^4\right)\,.
\label{eq:DeltaR_gen}
\ee
which is the usual approximation.
 
When $\Delta R$ is not small with respect to one,
Eq.~\eqref{eq:DeltaR_gen} is not valid anymore.  One can rather
consider the particular cases $\Delta y \ll \Delta \phi \approx \Delta
R$ and $\Delta \phi \ll \Delta y \approx \Delta R$, which give
respectively
\be
\Delta R = 
2\,\arctan\left(\frac{m}{|\mathbf{p}_T|}\right)
+O(\Delta y^2)\,,
\label{eq:DRphi}
\ee\be
\Delta R = 
2\, \operatorname{arcsinh} \left(\frac{m}{|\mathbf{p}_T|}\right)
+O(\Delta \phi^2)\,.
\label{eq:DRy}
\ee
These approximations will be used in our approach to jet merging.

Finally, it is also necessary to consider configurations giving an
upper bound on $\Delta R$.  These arise from decays with asymmetric
transverse momentum.  A sensible condition on the asymmetry is the one
given by the experimental jet definition.  We use the standard
asymmetry measure \cite{BDRS}
\be
\tau=\frac{\min(|\mathbf{p}_{T\,1}|^2, |\mathbf{p}_{T\,2}|^2)}{m^2}\Delta R^2\,. \label{eq:tau}
\ee
Below a threshold $\tau_{\rm cut}$, the jet is considered to be too
asymmetric to be likely to arise from the decay of a massive particle.
We write $|\mathbf{p}_{T,2}|=a|\mathbf{p}_{T,1}|$, choosing $a>1$
without loss of generality.  Assuming $\Delta \phi\ll \Delta y$, one
gets
\be
\Delta R = 2\, {\rm arcsinh}\left( \frac{m}{2|\mathbf{p}_{T}|}\frac{1+a}{\sqrt{a}} \right) +O\left(\Delta\phi^2\right)\,. \label{eq:DRasym}
\ee
Combining the asymmetry threshold $\tau\equiv\tau_{\rm cut}$ and
Eqs.~\eqref{eq:tau}, \eqref{eq:DRasym}, one gets the threshold value
$a_{\rm cut}$.  This is $a_{\rm cut}=1/\tau_{\rm cut}$ in the small
angle limit, \ie~$a\gg 1$, and has to be obtained numerically if this
condition is not fulfilled.  This provides the upper bound $\Delta
R_{\rm cut}=\Delta R (a=a_{\rm cut})$ which is used in
Sec.~\ref{se:top_partners}.
 
We can now use these expressions to estimate the region where $t'$ fat
jets are likely to occur.  Clearly, $t'$ decays tend to be more
collimated at high $\mathbf{p}_T$.  However, a subtlety is that the
subsequent $t,h,Z$ and $W$ jets should also get more collimated as
they inherit a higher $\mathbf{p}_T$ from the mother particle.  Our
strategy is to look for the most favorable phase space configuration.
If this configuration does not lead to jet merging, then the $t'$
decays are resolved over the whole phase space.  This most favorable
configuration is for a $t'$ decaying at minimal $\Delta R$ and at zero
rapidity, and for daughter particles decaying at maximal $\Delta R$ as
determined by the asymmetry cut.  The opening angle for the $t'$ decay
is given by Eqs.~\eqref{eq:DRphi}, \eqref{eq:DRy}.\footnote{ These two
limit cases lead respectively to daughters with
$|\mathbf{p}_T|_{t'}/2$ and $\sqrt{m^2+|\mathbf{p}_T|_{t'}}/2$.} The
daughters (\ie~$t$, $h$, $W$, $Z$) decay asymmetrically with $\Delta
R$ given by Eq.~\eqref{eq:DRasym}, using the standard cut $\tau_{\rm
cut}=0.09$.

The condition for two jets $1,2$ arising from a same vertex to be
resolved is \be\Delta R_{12} \leq R_1+ R_2\,.\label{eq:R_cond}\ee When
this condition is not fulfilled, the radius of the single jet formed
by the two merging jets is
\be
 R = \max\left[R_1,R_2, \frac{\Delta R_{12} + R_1+ R_2}{2} \right]\,.\label{eq:R_a}
\ee
Applied to the $t'$ decay, the condition Eq.~\eqref{eq:R_cond}
determines whether the $t'$ decay products are resolved.  The radius
of the $t,h,Z,W$ jets is described by Eq.~\eqref{eq:R_a}.\footnote{In
the case of the top decay, the subsequent $W$ decays asymmetrically
using again Eq.~\eqref{eq:DRasym}.  The $|\mathbf{p}_T|$ of the $W$
satisfies $|\mathbf{p}_T|_W=a_{\rm cut}|\mathbf{p}_T|_t$.  } Finally,
we also need the $t'$ transverse momentum at zero rapidity.  This is a
function of the global Higgs and $t'$ masses, given by
\begin{eqnarray}
|\mathbf{p}_T|_{t'} &=& \frac{1}{2}\sqrt{m_\phi^2-4m_{t'}^2}\,\quad \textrm{for}\quad  \phi\rightarrow t'\bar t\, (t\bar t')\,, \\
|\mathbf{p}_T|_{t'} &=& \frac{1}{2}\left(m_\phi-\frac{m_{t'}^2}{m_\phi}\right)\,\quad \textrm{for}\quad  \phi\rightarrow t'\bar t'\,,
\end{eqnarray}
when the top mass is neglected.

Putting all these pieces together provides a region of the parameter
space where $t'$-jets can potentially occur.  This region is displayed
for every decay in Fig.~\ref{fig:boost_TpTpbar}.  For the asymmetry
criteria $\tau_{\rm cut}=0.09$, it turns out this matches roughly the
region with a fraction of $2-5\%$ of $t'$-jets.  The region obtained
in case of azimuthal $t'$ decay configuration Eq.~\eqref{eq:DRphi}
turns out to be larger than for polar decay Eq.~\eqref{eq:DRy}, so
that we display only the former.

\bibliographystyle{JHEP} 

\bibliography{SO5_biblio}

\end{document}